\documentclass[sigconf,9pt]{acmart}
\AtBeginDocument{%
  \providecommand\BibTeX{{%
    Bib\TeX}}}

\copyrightyear{2023}
\acmYear{2023}
\setcopyright{acmlicensed}\acmConference[ICS '23]{2023 International Conference on Supercomputing}{June 21--23, 2023}{Orlando, FL, USA}
\acmBooktitle{2023 International Conference on Supercomputing (ICS '23), June 21--23, 2023, Orlando, FL, USA}
\acmPrice{15.00}
\acmDOI{10.1145/3577193.3593705}
\acmISBN{979-8-4007-0056-9/23/06}

\usepackage{algorithmic}
\usepackage{graphicx}
\usepackage{textcomp}
\usepackage{xcolor}
\usepackage{url}
\usepackage{hyperref}
\usepackage{enumitem}
\usepackage{soul}
\usepackage[outline]{contour}
\usepackage{hhline}
\usepackage{soul}

\setlist[itemize]{leftmargin=3.0mm}
\def\BibTeX{{\rm B\kern-.05em{\sc i\kern-.025em b}\kern-.08em
    T\kern-.1667em\lower.7ex\hbox{E}\kern-.125emX}}

\usepackage[skip=0pt]{caption}
\usepackage{colortbl}
\newcommand{\wahib}[1] {{\color{blue}{Wahib}}: {\color{blue}{#1}}}
\newcommand{\lingqi}[1] {{\color{red}{Lingqi}}: {\color{red}{#1}}}
\newcommand{\chen}[1] {{\color{gray}{Chen}}: {\color{green}{#1}}}

\usepackage{listings}
\usepackage{color}
\definecolor{codegreen}{rgb}{0,0.6,0}
\definecolor{codegray}{rgb}{0.5,0.5,0.5}
\definecolor{codepurple}{rgb}{0.58,0,0.82}
\definecolor{backcolour}{rgb}{0.95,0.95,0.92}
 
\usepackage{caption}
\lstdefinestyle{mystyle}{
    backgroundcolor=\color{white},   
    commentstyle=\color{codegreen},
    keywordstyle=\color{magenta},
    numberstyle=\tiny\color{codegray},
    stringstyle=\color{codepurple},
    basicstyle=\footnotesize,
    breakatwhitespace=false,         
    breaklines=true,                 
    captionpos=b,                    
    keepspaces=true,                 
    numbers=left,                    
    numbersep=5pt,                  
    showspaces=false,                
    showstringspaces=false,
    showtabs=false, 
    tabsize=2
}
 
\usepackage{comment}
\usepackage{adjustbox}
\DeclareRobustCommand*\circled[1]{\tikz[baseline=(char.base)]{
            \node[shape=circle,fill,inner sep=1.3pt] (char) {\textcolor{white}{#1}};}}
\newcommand{\tbhline}{\noalign{\hrule height 1pt}}

\usepackage[switch]{lineno} 

\newcommand{\revision}[1]{{\color{black}{#1}}}

\newcolumntype{|}{!{\vrule width 0.5pt}}
\newcolumntype{?}{!{\vrule width 1pt}}
\newcolumntype{^}{!{\vrule width 1.2pt}}

\makeatletter
  \def\title@font{\Large}
  \let\ltx@maketitle\@maketitle
  \def\@maketitle{\bgroup%
    \let\ltx@title\@title%
    \def\@title{\resizebox{\textwidth}{!}{%
      \mbox{\title@font\ltx@title}%
    }}%
    \ltx@maketitle%
  \egroup}
\makeatother
\usepackage{multirow}
\usepackage{tablefootnote}
\usepackage{todonotes}
\usepackage{subcaption}

\definecolor{Gray}{gray}{0.9}

\DeclareRobustCommand{\hllightgray}[1]{{\sethlcolor{lightgray}\hl{#1}}}

\makeatletter
\DeclareUrlCommand\ULurl@@{%
  }
\def\ULurl@#1{\hyper@linkurl{\ULurl@@{#1}}{#1}}
\DeclareRobustCommand*\ULurl{\hyper@normalise\ULurl@}
\makeatother
\begin{document}

%%
%% The "title" command has an optional parameter,
%% allowing the author to define a "short title" to be used in page headers.
% \title{Persistent Kernels for Iterative Memory-bound GPU Applications}
%\title{Optimize Iterative Memory-bound Applications Integrally with Persistent Kernels}
% \title{Breaking the Memory Bottleneck for Iterative Memory-bound Applications Via Persistent Kernels}
\title{PERKS: a Locality-Optimized Execution Model for Iterative Memory-bound GPU Applications}%\vspace{-10pt}
% \title{Persistent Kernels for Iterative Memory-bound GPU Applications}

%%
%% The "author" command and its associated commands are used to define
%% the authors and their affiliations.
%% Of note is the shared affiliation of the first two authors, and the
%% "authornote" and "authornotemark" commands
%% used to denote shared contribution to the research.

% \settopmatter{authorsperrow=4}
\author{Lingqi Zhang}
\affiliation{%
  \institution{Tokyo Institute of Technology}
  % \cite{Tokyo}
  \country{Japan}
}
\additionalaffiliation
{
    \institution{National Institute of Advanced Industrial Science and Technology}
  % \city{Tokyo}
  \country{Japan}
}
\email{zhang.l.ai@m.titech.ac.jp}

\author{Mohamed Wahib}

\authornote{Corresponding authors}
\affiliation{%
  \institution{RIKEN Center for Computational Science}
  % \cite{Tokyo}
  \country{Japan}
}
\email{mohamed.attia@riken.jp}
\author{Peng Chen}
\authornotemark[2]
\affiliation{%
  \institution{National Institute of Advanced Industrial Science and Technology}
  \country{Japan}
}
\additionalaffiliation
{
    \institution{RIKEN Center for Computational Science}
  % \city{Tokyo}
  \country{Japan}
}
\email{chin.hou@aist.go.jp }
\author{Jintao Meng}
\affiliation{%
  % \institution{Chinese Academy of Sciences}
  \institution{Shenzhen Institutes of Advanced Technology}
  \country{China}
}
\email{jt.meng@siat.ac.cn}
\author{Xiao Wang}
\authornotemark[2]
\affiliation{%
  \institution{Oak Ridge National Laboratory}
  % \institution{ORNL}
  \country{USA}
}
\email{wangx2@ornl.gov}
\author{Toshio Endo}
\affiliation{%
  \institution{Tokyo Institute of Technology}
  \country{Japan}
}
\email{endo@is.titech.ac.jp}
\author{Satoshi Matsuoka}
% \authornotemark[1]
\affiliation{%
  \institution{RIKEN Center for Computational Science}
  \country{Japan}
}
\email{matsu@acm.org}
\additionalaffiliation
{
    \institution{Tokyo Institute of Technology}
  % \city{Tokyo}
  \country{Japan}
}
\renewcommand{\shortauthors}{Lingqi Z. et al.}

\begin{abstract}

Iterative memory-bound solvers commonly occur in HPC codes. Typical GPU implementations have a loop on the host side that invokes the GPU kernel as much as time/algorithm steps there are. The termination of each kernel implicitly acts the barrier required after advancing the solution every time step. We propose an execution model for running memory-bound iterative GPU kernels: PERsistent KernelS (PERKS). In this model, the time loop is moved inside persistent kernel, and device-wide barriers are used for synchronization. We then reduce the traffic to device memory by caching subset of the output in each time step in the unused registers and shared memory. PERKS can be generalized to any iterative solver: they largely independent of the solver's implementation. We explain the design principle of PERKS and demonstrate effectiveness of PERKS for a wide range of iterative 2D/3D stencil benchmarks (geomean speedup of $2.12$x for 2D stencils and $1.24$x for 3D stencils over state-of-art libraries), and a Krylov subspace conjugate gradient solver (geomean speedup of $4.86$x in smaller SpMV datasets from SuiteSparse and $1.43$x in larger SpMV datasets over a state-of-art library). All PERKS-based implementations available at: \ULurl{https://github.com/neozhang307/PERKS}.

% \ULurl{https://anonymous.4open.science/r/PERKS-F619/}

\end{abstract}

%(geomean speedup of $2.41$x for 2D stencils and $2.28$x for 3D stencils)

%%f
%% The code below is generated by the tool at http://dl.acm.org/ccs.cfm.
%% Please copy and paste the code instead of the example below.
%%

% \begin{CCSXML}
% <ccs2012>
%  <concept>
%   <concept_id>10010520.10010553.10010562</concept_id>
%   <concept_desc>Computer systems organization~Embedded systems</concept_desc>
%   <concept_significance>500</concept_significance>
%  </concept>
%  <concept>
%   <concept_id>10010520.10010575.10010755</concept_id>
%   <concept_desc>Computer systems organization~Redundancy</concept_desc>
%   <concept_significance>300</concept_significance>
%  </concept>
%  <concept>
%   <concept_id>10010520.10010553.10010554</concept_id>
%   <concept_desc>Computer systems organization~Robotics</concept_desc>
%   <concept_significance>100</concept_significance>
%  </concept>
%  <concept>
%   <concept_id>10003033.10003083.10003095</concept_id>
%   <concept_desc>Networks~Network reliability</concept_desc>
%   <concept_significance>100</concept_significance>
%  </concept>
% </ccs2012>
% \end{CCSXML}

% \ccsdesc[500]{Computer systems organization~Embedded systems}
% \ccsdesc[300]{Computer systems organization~Redundancy}
% \ccsdesc{Computer systems organization~Robotics}
% \ccsdesc[100]{Networks~Network reliability}

%%
%% Keywords. The author(s) should pick words that accurately describe
%% the work being presented. Separate the keywords with commas.
%%\keywords{Persistent GPU Kernels, Iterative Solvers, Stencil, Conjugate Gradient Solvers}


\maketitle

\section{Introduction}\label{sec:intro}

GPUs are becoming increasingly prevalent in HPC systems. More than half the systems on the Top500~\cite{top500} list include discrete GPUs and seven of the systems in the top ten are GPU-accelerated (November 2022 list). As a result, extensive efforts \revision{goes} into optimizing iterative methods for GPUs, for instance: iterative stencils~\cite{meng2011performance,8820786,DBLP:conf/cgo/MatsumuraZWEM20,chen2019versatile} used widely in numerical solvers for PDEs, iterative stationary methods for solving systems of linear equations (ex: Jacobi ~\cite{ahamed2017efficient,kochurov2015gpu}, Gauss–Seidel method~\cite{courtecuisse2009parallel,fratarcangeli2016vivace,kochurov2015gpu}), iterative Krylov subspace methods for solving systems of linear equations (ex: conjugate gradient~\cite{phillips2014cuda,anzt2020ginkgo}, BiCG\cite{anzt2020ginkgo,aliaga2015systematic}, and GMRES\cite{anzt2020ginkgo,couturier2012sparse}). 

%but i can ping you on Line around sunday when i knnow better

%\wahib{@Lingqi: 1 paragraph on the challenges}
Although the device memory bandwidth of GPUs has been increasing from generation to generation, the gap between compute and memory is widening. Given that iterative stencils and implicit solvers typically have low arithmetic intensity~\cite{DBLP:conf/cgo/MatsumuraZWEM20}, significant efforts \revision{goes} into optimizing them for data locality. These included moving the bottleneck from device memory to on-chip scratchpad memory~\cite{maruyama2014optimizing} or cache~\cite{malas2015multicore}, or further pushing the bottleneck to the register files~\cite{chen2019versatile,zhao2019exploiting}. Those efforts become increasingly effective since the aggregate volume of register files and scratchpad memory capacity are increasing with newer generations of GPUs~\cite{jia2019dissecting}. In iterative solvers, due to spatial dependencies, a barrier is typically required at the end of each time step (or several time steps when doing temporal blocking~\cite{DBLP:conf/cgo/MatsumuraZWEM20}). That is to assure that advancing the solution in time step $k$ would only start after all threads finish advancing the solution in time step $k-1$. Invoking the kernels from the host side in each time step acts as an implicit barrier, where the kernel invocation in time step $k$ would happen after all threads of the kernel invocation at time step $k-1$ have finished execution. In-between kernel invocations, data stored in registers and scratchpad memory would be wiped out, and the next kernel invocation would start by reading its input from the device memory. 

%Memory-bound kernels have regular memory access patterns, yet sometimes hard for the compiler to automatically detect the potential of memory reuse pattern. 

% Yet its volume is stile relative small compared with the increasing in global memory, and the increasing in problem size. The aim of this work is to 

%\wahib{@Lingqi: 1 paragraph to introduce the proposed solution}

One opportunity to improve the data locality is to extend the lifetime of the kernel across the time steps and take advantage of the large volume of register files and scratchpad memory to reduce traffic to the device memory. In this paper, we propose a generic model for running iterative solvers on GPUs to improve data locality. PERsistent KernelS (PERKS)\footnote{In this paper, we use PERKS, interchangeably, to refer to our proposed model and, as an abbreviation of PERsistent KernelS} have the time loop inside them, instead of the host, and use the recently supported device-wide barriers (in CUDA) for synchronization. Next, we identify the cache-able data in the solver: the data that is the output of time step $k-1$ and input to time step $k$, as well as the repeatedly loaded constant data. Finally, we use either the scratchpad memory or registers (or both) to cache the data, and reduce the traffic to device memory. %Additionally, we explore a further improvement on PERKS, namely tiled PERKS, in which we tile the domain and serialize the tiles. PERKS are tiled to tile sizes that could be entirely cachable, and hence could be computed with very limited amount of traffic to the device memory. This approach is particularly effective in 2D domains. 

The basic concept and implementation of PERKS is relatively simple, which we argue is essential for encouraging scientists and engineers to adopt PERKS in their iterative solvers implemented for GPUs, and other architectures as well. That being said, a challenging aspect that we address in this paper is a detailed analysis of how and why PERKS is effective. The analysis requires an understanding of the effect of concurrency on performance. More particularly, to gain a deep understanding of why PERKS are effective and the limitations of architectural features, we study the effect of pressure on resources (particularly registers and shared memory). On top of that, we examine the effect of reducing the device occupancy while maintaining high enough concurrency to saturate the device.

%Fourth, to make PERKS a practical scheme, a simple programming construct should be exposed for programmers to port their solvers to be PERKS with minimal effort. 
%Second, we need to analyze and pick the ideal geometric scheme to traverse the domain of input array(s) at each step to improve the locality and increase the volume of cached data when in transition between time steps.

It is important to note that PERKS are orthogonal to temporal blocking optimizations. Temporal blocking relies on combining multiple consecutive iterations of the time loop to reduce the memory transactions between them. The dependency along the time dimension is resolved by either: a) redundantly loading and computing cells from adjacent blocks, which limits the effectiveness of temporal blocking to low degrees of temporal blocking~\cite{10.1145/2304576.2304619,10.1145/1542275.1542313,10.1145/2830018.2830025}, or b) using tiling methods of complex geometry (e.g. trapezoidal and  hexagonal tiling) along the time dimension and restrict the parallelism due to the dependency between neighboring blocks~\cite{7582549,10.1145/2458523.2458526,MURANUSHI20151303}. In contrast, the execution model of PERKS does not necessitate the resolution of the dependency along the time dimension since PERKS include an explicit barrier after each time step. This means the PERKS model can be generalized to any iterative solver and can be used on top of any version of the solver. In other words, iterative kernels written as PERKS do not compete with optimized versions of those iterative kernels. For instance, a stencil PERKS does not compete with kernels applying aggressive locality optimizations~\cite{chen2019versatile,chen2018efficient} and out-of-core optimizations~\cite{sabet2020subway,endo2018applying,wahib2020scaling}; the performance gain from PERKS is added to the performance gain from whatever stencil optimizations are used in the kernel. \textbf{As a matter of fact, the more optimized the kernel before it is ported to the PERKS execution model, the higher the speedup that would be gained by PERKS. That is since optimizations to the kernel proportionally increases the overhead of data storing and loading in between iterations, which PERKS aim to reduce.}

    on A100 and V100. PERKS-based conjugate gradient achieves a geometric mean speedup of \revision{$2.48$x} in comparison to the highly GPU-optimized production library Ginkgo~\cite{anzt2020ginkgo} for SpMV datasets in SuiteSparse. For smaller datasets, the speedup goes up to \revision{$2.86$x}. %The source code for all PERKS-based implementations in this paper is available at the following anonymized link:~\href{http://shorturl.at/cdjmX}{\color{blue}{http://shorturl.at/cdjmX}}.%, in addition to the sample code listed in the supplementary material.
    % \item Our PERKS-based implementation achieves geometric means speedups of $2.35$x for 2D stencils and $1.53$x for 3D stencils using highly optimized baselines comparable to state-of-the-art 2D/3D stencil implementations, with A100 and V100. PERKS-based conjugate gradient achieves a geometric mean speedup of $2.47$x in comparison to the highly GPU-optimized production library Ginkgo~\cite{anzt2020ginkgo} for SpMV datasets in SuiteSparse. For smaller datasets, the speedup goes up to $4.60$x. The source code of all PERKS-based implementations in this paper is available at the following anonymized link:~\href{http://shorturl.at/cdjmX}{\color{blue}{http://shorturl.at/cdjmX}}, in addition to sample code listed in the supplementary material.
\end{itemize}

% \lingqi{
% % \color{red}
The rest of this paper is organized as follows:
Section~\ref{sec:background}, presents the background and motivation.
Section~\ref{sec:method} presents the overview of the execution model PERKS.
Section~\ref{sec:imp} shows the implementation of PERKS. 
% \lingqi{We introduce the implementation as an end-user perspective. We also elaborate in details how we implement PERKS for stencil kernels and conjugate gradient solver kernels.}
In Section~\ref{sec:simplepermodel}, we analyze the performance consideration of PERKS. 
Section~\ref{sec:eval} displays the evaluated result.
In Section~\ref{sec:related}, we elaborate on the related work.
Finally, Section~\ref{sec:conclusion} concludes the paper.
% }
\section{Background and Motivation}\label{sec:background}

\subsection{CUDA Programming Model}\label{sec:backsync}
CUDA's programming model~\cite{nvidia2019programming} includes: \textit{threads}, the basic execution unit (32 threads are executed together as a \textit{warp}); \textit{Thread block (TB)}, usually composed of hundreds of threads; \textit{grid}, usually composed of tens of thread blocks. 

\subsubsection{\revision{GPU Memory Hierarchy:}} On-chip memory in a streaming multiprocessor (SM) includes: shared memory (scratchpad memory), L1 cache, and register file (RF) and. Off-chip memory includes \revision{global memory (device memory)} and L2 cache. Data in global memory can reside for the entirety of the program, while data in on-chip memory has the lifetime of a kernel. The shared memory is shared among all threads inside a thread block. \revision{We summarize the relevant memory features in Table~\ref{tab:gpu_loadsmltc}.}

\begin{table}[t]
    \centering
    \caption{Relevant Features of the latest Nvidia GPUs\cite{nvidiaa100}}% SM is Steam Multiprocessor}
    % \resizebox{\linewidth}{!}
    {
    \small
    \begin{tabular}{?l?c|c|c?}\tbhline
    \textbf{Feature}    & \textbf{P100} & \textbf{V100}  & \textbf{A100}  \\\hline
    \textbf{Shared memory}~\tablefootnote{\emph{Shared memory} is a configurable portion of L1 cache that can be used as a user-managed scratchpad memory} 
    & 3.5 MB & 7.5 MB & 17.29 MB \\
    \textbf{Register files} &14 MB &20MB&27MB\\
    \textbf{L2 cache }    & 4 MB & 6 MB  & 40 MB  \\
    \textbf{Memory bandwidth}    & 720 GB/s & 900 GB/s  & 1555 GB/s  \\\hline
    % \textbf{Time}    &   &  & \\
    % \textbf{(Fill $50\%$ of register files)}    &  10.19 ns & 11.65 ns & 9.10 ns \\\hline
    \end{tabular}
    \label{tab:gpu_loadsmltc}
    }
\end{table}

\subsubsection{GPU Device-wide Synchronization:} Synchronization in GPUs was limited to groups of threads: thread blocks in CUDA (or a work group in OpenCL). Starting from CUDA 9.0, Nvidia introduced cooperative group APIs~\cite{nvidia2021api} that include an API for device-wide synchronization. Before introducing grid-level synchronization, the typical way to introduce device-wide synchronization was to launch sequences of kernels in a single CUDA stream. Zhang et al.~\cite{zhang2020study} conducted a comprehensive study to compare the performance of both methods. The result shows that the latency difference between explicit device-wide synchronization versus implicit synchronization (via repetitive launching of kernels) is negligible in most kernels.

\subsection{Iterative Algorithms}\label{sec:iteralrithm}
In iterative algorithms, the output of time step $k$ is the input of time step $k+1$. Iterative methods can be expressed as: 

\begin{equation}\footnotesize
    \begin{aligned}
    x^{k+1}= F(x^k)
    \end{aligned}
    \label{eqt:iterative}
\end{equation}

When the domain is mapped out to processing elements, there are two points to consider:
\begin{itemize}[leftmargin=2.5mm]
\item Spatial dependency necessitates synchronization between time steps, or else advancing the solution in the following time step might use data that has not yet been updated in the previous time step.  
\item In time step {\footnotesize{$k+1$}}, each thread or thread block needs input from the output of itself in time step {\footnotesize$k$} (i.e. temporal dependency). This gives the opportunity for caching data between steps to reduce device memory traffic. 
\end{itemize}
In the following sections, we briefly introduce iterative stencils and Krylov subspace methods. Throughout the paper, we use them as motivation examples, and we use them to report the effectiveness of our proposed methods, given their importance in HPC scientific and engineering codes.

\subsubsection{Iterative Stencils}
Iterative stencils are widely used in HPC. According to Bastian et al.~\cite{hagedorn2018high}, stencil applications represent 49\% of workloads in a wide range of HPC centers. Take 2D Jacobian 5-point stencil (2d5pt) as an example:%\todo{CHANGE every example in this paper to 2d5pt}:
\begin{equation}\footnotesize
    \begin{aligned}
    x(i,j)^{k+1} =& N*x(i,j+1)^k + S*x(i,j-1)^k + \\
    &C*x(i,j)^k + W*x(i-1,j)^k + E*x(i+1,j)^k
    \end{aligned}
    \label{eqt:iterativeStencil}
\end{equation}

% \revision{2d5pt stencil is also used a}

Computation of each point at time step {\footnotesize$k+1$} requires the values of the point itself and its four neighboring points at time step $k$. 

Two blocking methods are widely used to optimize iterative stencils for data locality: \textit{Spatial Blocking}~\cite{irigoin1988supernode,Wolfe:1989:MIS:76263.76337} and \textit{Temporal Blocking}~\cite{DBLP:conf/cgo/MatsumuraZWEM20,rawat2016effective,rawat2018domain}. 

In spatial blocking on GPUs, we split the whole domain into sub-domains, where each thread block can load its sub-domain to the shared memory to improve the data reuse. In the meantime, we require redundant data accesses at the boundary of the thread block to data designated for adjacent thread blocks. 

In iterative stencils, each time step depends on the result of the previous time step. One could advance the solution by combining several time steps. The temporal dependency, in this case, is resolved by using a number of halo layers that match the number of combined steps. The amount of data that can be computed depends on the stencil radius ($rad$) and the number of time steps that are combined ($b_t$). In overlapped temporal tiling~\cite{Holewinski:2012:HCG:2304576.2304619,Krishnamoorthy:2007:EAP:1250734.1250761,Meng:2009:PMA:1542275.1542313}, this region can be represented as {$2 \times b_t\times rad$} ($halo$ region). Methods based on this kind of blocking are called \textit{overlapped temporal blocking} schemes. Overlapped temporal blocking introduced the overhead of redundant computation that  wavefront~\cite{malas2015multicore,wellein2009efficient,belviranli2015peerwave} is aimed to alleviate.

\subsubsection{Krylov Subspace Methods}

Krylov methods are widely used for large sparse (and dense) linear systems of equations arising in solvers of Partial Differential Equations (PDEs)~\cite{anzt2017preconditioned,frommer2017block,pearson2020preconditioners}. Krylov subspace methods can be described as:
\begin{equation}\footnotesize
    \begin{aligned}
    \kappa _r (A,b)=span\{b,Ab,A^{2},...,A^{r-1}b\}
    \end{aligned}
    \label{eqt:kryslov}
\end{equation}

Assuming that A is an invertible matrix, it is possible to compute {\footnotesize$x=A^{-1}b$} (or solve {\footnotesize$Ax=b$}) by searching the Krylov subspace without directly computing {\footnotesize$A^{-1}$}. Searching the Krylov subspace is a sequence of matrix-vector multiplications, where at each step the approximation of the solution vector $x$ is updated proportionally to the residual error (vector $r$) from the previous time step.

Conjugate gradient is a main solver in the family of Krylov subspace methods. It is mainly used to solve systems of linear equations for symmetric and positive-definite matrices.

\begin{comment}
\begin{algorithm}[t]
% \LinesNumbered
% \begin{algorithmic}[1]
\SetAlgoLined
% \KwResult{Write here the result }
 // Given A, a symmetric positive-definite matrix\;
 // We solve Ax=b\;
 $x_0=0$;
 $r_0=b$;
 $p_0=b$\;
 \While{$k<k_{max}$}{
 $\alpha_k=\frac{<r_k,r_k>}{p_k,Ap_k}$\;
 $x_{k+1}=x_k+\alpha_kp_k$\;
 $r_{k+1}=r_k-\alpha_kAp_k$\;
 $\beta_k=\frac{<r_{k+1},r_{k+1}>}{<r_k,r_k>}$\;
 $p_{k+1}=r_{k+1}+\beta_kp_k$\;
    \If{$<r_{k+1},r_{k+1}>>tol \times tol$}{
    break\;
    }
 }
 
 \caption{\label{algorithm:cg}Conjugate Gradient Solver}
% \end{algorithmic}
\end{algorithm}
\end{comment}

%% COMMENTED THE FIGURE OUT
\begin{comment}
\begin{figure}[t]
\includegraphics[width=0.45\textwidth]{FIG/Memory_Hierarchy.pdf}
\caption{\label{fig:memory_hiearchy} The relationship between CUDA Programming Model and GPU memory hierarchy. Italic format in color filled box marks CUDA Programming Model. Bank filled box marks represents memory hierarchy.

}
\end{figure}
\end{comment}

\subsection{Motivational Example}

\begin{figure}[t]
\centering
\centering
\includegraphics[width=\linewidth]{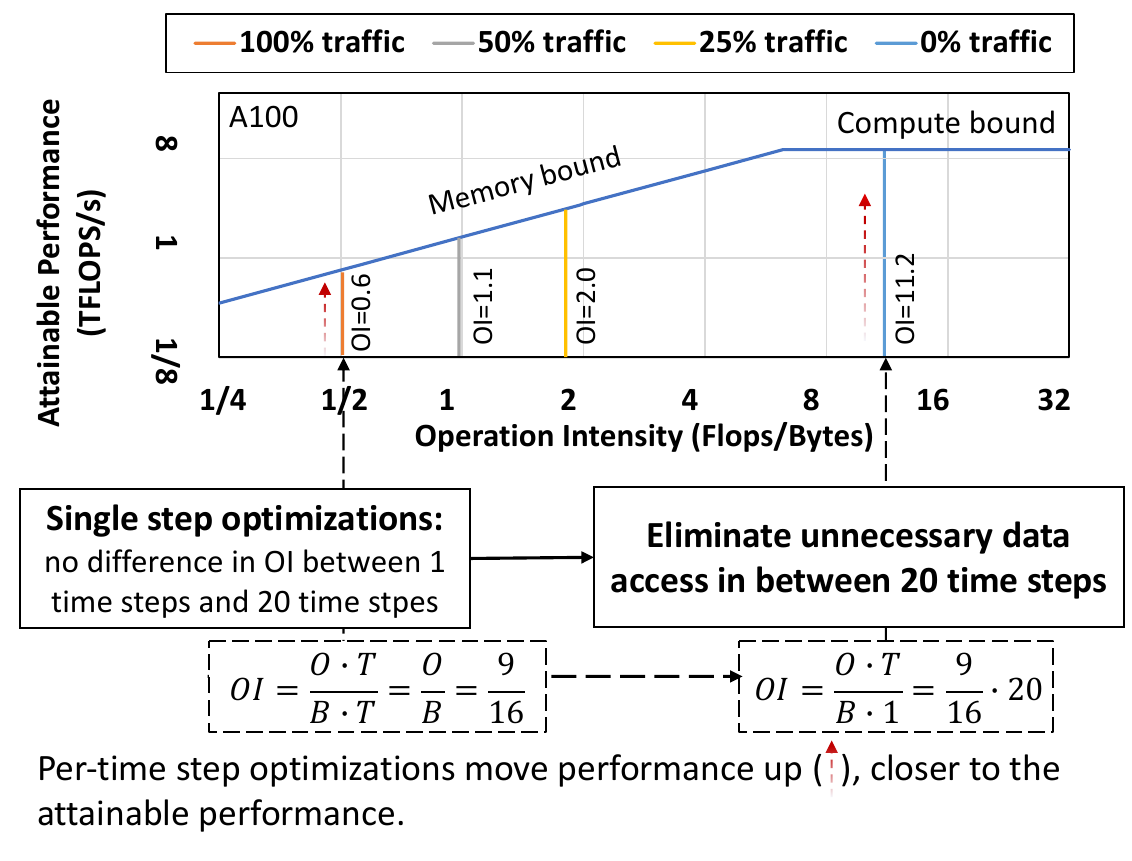}
\caption{The roofline model of a 2D 5-point Jacobian stencil kernel (dp), running $T=20$ time steps, with a domain size of $3072^2$ on A100 GPU. \revision{Per-time step optimizations only improve the iterative stencil kernel to get closer to the attainable performance.} Reducing memory traffic between time steps can increase the \revision{attainable} performance by increasing the aggregate operation intensity (OI) over all time steps. We \revision{also} plot different operational intensities for a version of PERKS that reduces the data traffic in-between 20 time steps to $50\%$, $25\%$, and $0\%$.}
\label{Fig:roofline}
\end{figure}
% Optimization per-time step uses shared memory to improve locality~\cite{maruyama2014optimizing}. 

\begin{figure}[t]
\centering
\centering
\includegraphics[width=\linewidth]{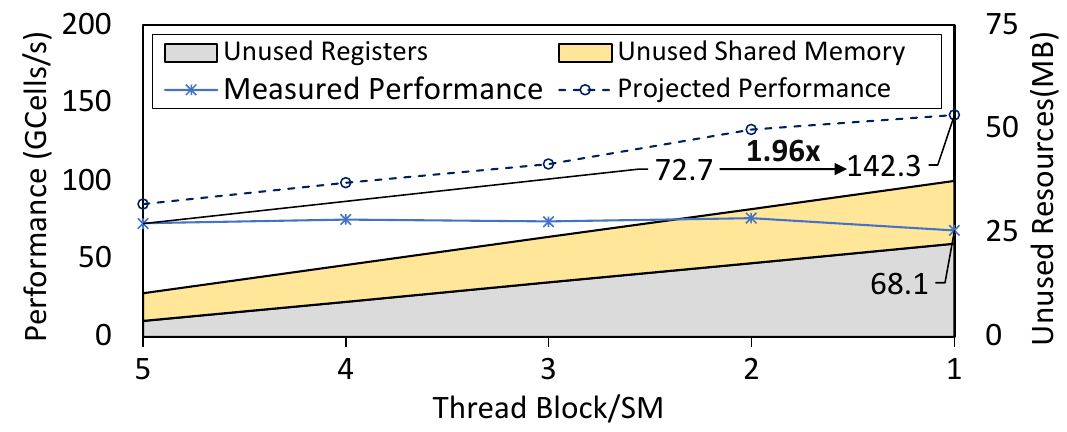}
\caption{Performance of a double precision 2D \revision{5-point} Jacobian stencil kernel ($3072^2$) for different Thread Blocks per Streaming Multiprocessor (TB/SM) on A100. Filled regions indicate unused resources. Using one TB/SM and using all unused resources for caching can theoretically provide $1.96$x speedup in this situation.} 
\label{Fig:showcase}
\end{figure}
%\revision{and we \textbf{aggressively} assume that any portion of memory traffic saved directly benefit performance.}
%\revision{We assume} that \revision{we can use} all unused resources to cache data. 
% \todo{Think a better way to present}
\begin{figure}[t]
{
% \frame
{\includegraphics[width=\linewidth]{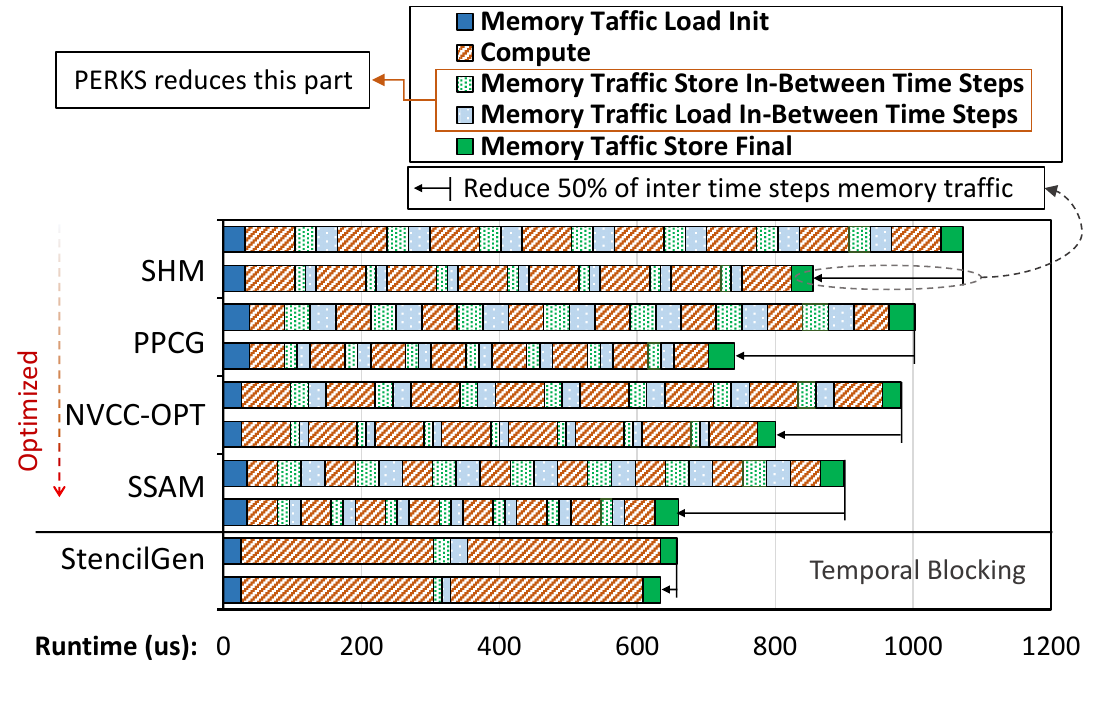}}}
\caption{Runtime (\revision{8} time steps) of double precision 2D \revision{5-point} Jacobian stencil ($3072^2$) with different state-of-the-art optimization A100 GPU.
PERKS aims to reduce the traffic to/from device memory in-between time steps. SHM~\cite{maruyama2014optimizing} uses shared memory. PPCG is a code auto generation tool~\cite{verdoolaege2013polyhedral}. NVCC-OPT relies on auto optimization provided by the latest NVCC compiler version. SSAM~\cite{chen2019versatile} uses register to improve locality~\cite{chen2018efficient,ifdk,dfbp}. Stencilgen~\cite{rawat2018domain} applies temporal blocking.
In the figure we also plot the total runtime of each implementation running 8 time steps, assuming we reduce $50$\% of the inter time step memory traffic.
The results show that the more optimized \revision{(i.e. fewer proportion is spent in compute)}, the more performance improvement we expect from caching.}
\label{Fig:showLatency}
\end{figure}

We use a motivational example of a double precision \revision{2D 5-point} Jacobian \revision{(2d5pt)} stencil to motivate implementing iterative solvers as PERKS \revision{(Readers can refer to Equation~\ref{eqt:iterativeStencil} and Listing~\ref{Fig:codeperk22d5pt} for details of the 2d5pt stencil)}. 
\circled{1} 
Why PERKS: Optimizations for iterative methods focus on a single step to speed up iterative solvers. Single-step optimizations move the performance of the kernel closer to the highest possible attainable performance on the roofline model, yet will not influence the operational intensity. As Figure~\ref{Fig:roofline} shows, the optimizations used for the 2D 5-point stencil move the performance vertically at the same operational intensity value of the kernel. Temporal blocking schemes can horizontally move the operational intensity, to the right side of the roofline. Yet resolving the neighborhood dependencies introduces redundancy~\cite{Holewinski:2012:HCG:2304576.2304619,Krishnamoorthy:2007:EAP:1250734.1250761,10.1145/2830018.2830025} or hard-to-parallel complex geometrical tile shapes~\cite{bondhugula2017diamond,10.1145/2458523.2458526,MURANUSHI20151303}, and can cause register pressure~\cite{DBLP:conf/cgo/MatsumuraZWEM20}. In PERKS,
% we batch a sequence of time steps together and remove
\revision{we reduce}
the unnecessary data access between time steps. The target data traffic to reduce is in-between time steps (i.e., outside the solver) and hence is not subject to the neighborhood dependency issue in temporal blocking schemes. Figure~\ref{Fig:roofline} demonstrates how this idea works for a real stencil benchmark running on an A100 GPU for 20 time steps. By caching more of the domain in-between time steps, the operational intensity moves more to the right side of the roofline to be compute-bound. This also demonstrates how PERKS is orthogonal to the per-time step optimizations; PERKS would improve the performance (by moving horizontally on the roofline) regardless of how optimized the baseline algorithm is at its operational intensity.
\circled{2} 
The prospect of PERKS: Latency across all operations/instructions in newer generation GPUs has been significantly dropping~\cite{DBLP:journals/corr/abs-1804-06826}. As a result, often fewer numbers of warps are enough for CUDA runtime to hide the latency effectively and hence maintain high performance at low occupancy~\cite{volkov2010better}. In Figure~\ref{Fig:showcase}, we vary the number of thread blocks per streaming multiprocessor (TB/SM) and plot its performance (left Y-axis). For each TB/SM configuration, we plot on the right Y-axis the unused resources (shared memory and registers). As the figure shows, \textbf{even when $\mathbf{TB/SM=5}$, more than $\mathbf{10.46}$ MB of shared memory and register files are not in use.} When $TB/SM$ decreases, the performance is slightly fluctuated (\revision{$72.74\rightarrow68.12$} GCells/s~\footnote{GCells/s denotes giga-cells updated per second.}) while the freed shared memory and registers gradually increase. By reducing TB/SM to its minimum while maintaining enough concurrency to sustain the performance level, the projection from performance gain when caching a subset of the results in unused resources can improve performance by more than \revision{$1.96$x}. 
\revision{Figure~\ref{Fig:showLatency} uses an alternative perspective, assuming that memory accessing and compute within a solver can not be overlapped. We profiled different 2d5pt state-of-the-art implementations.} As Figure~\ref{Fig:showLatency} shows, the compute part decreases as the more optimized the stencil implementation is. The prospect of PERKS is to reduce this data movement time that dominates the runtime in highly optimized stencil implementations. Note that while temporal-blocking schemes do also reduce the data movement to some extent, they cannot be generalized to all iterative solvers. Additionally, resolving temporal and spatial dependency adds compute overhead and can also lead to increased register pressure that limits the degree of temporal blocking on GPUs~\cite{DBLP:conf/cgo/MatsumuraZWEM20}.

% \section{PERKS: Persistent Kernels to Improve Data Locality}\label{sec:method}
% \section{PERKS: Persistent Kernels to Improve Locality}\label{sec:method}
\section{PERKS: Persistent Kernels to Improve Locality}\label{sec:method}
%%\section{PERKS: Persistent Kernels to Improve Locality}\label{sec:method}
\begin{figure}[t]
\centering
\includegraphics[width=\linewidth]{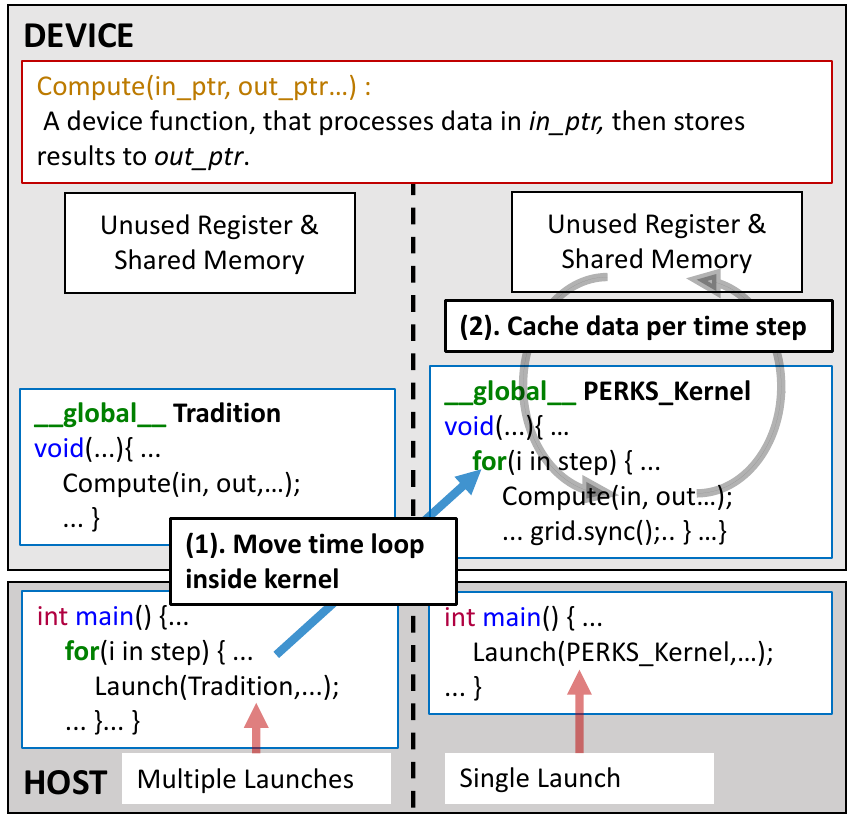}
\caption{\label{fig:exemodel} Changing a traditional iterative CUDA kernel (time loop on the host) to PERKS: 1) move the time loop from the host code to the kernel code and use grid synchronization between time steps. 2) Cache data between time loops on the unused shared memory and register files. The compute portion of the kernels does not notably change, i.e., no need to change the original algorithm when using PERKS.
}
\end{figure}
PERsistent Kernels (PERKS) is a generic \revision{execution model} for running iterative solvers on GPUs to improve data locality by taking advantage of the large capacity of register files and shared memory. As Figure~\ref{fig:exemodel} illustrates, in PERKS, we move the time stepping loop from the host to the device, and use CUDA's grid synchronization API as a device-wide barrier at each time step. \revision{This way, we expose the temporal data locality across time steps to thread blocks. }We then use the register files and shared memory to reduce traffic to the global memory by caching the domain (or a subset from it) in-between time steps.

\subsection{Assumptions and Limitations}
\label{sec:assumptions}
The techniques discussed in this paper are based on the following assumptions about the applications. 

\textbf{Target Applications:} In this paper, we target iterative kernels that are bounded by memory bandwidth. Although execution in a PERKS fashion makes no assumptions on the underlying implementation, optimal PERKS performance can sometimes require minor adaptations to the kernel. The changes, for instance, can be as simple as changing the thread block and grid sizes to reduce over-subscription, or more elaborate as favoring a specific SpMV method from the space of SpMV methods in the case of the conjugate gradient (more details in Section~\ref{sec:impCg}). Finally, despite not reporting results for compute-bound iterative kernels, it is important to note that compute-bound iterative kernels could potentially also benefit from becoming PERKS, if the kernel generates memory traffic in-between iterations that CUDA runtime can not effectively overlap with computation. 

%%\textbf{Impact on Optimized Kernels:} It is crucial to note that PERKS is orthogonal to the optimization level applied to the compute part of the kernel. As a matter of fact, the more optimized the baseline kernel, the more performance improvement we expect from PERKS. Because optimizations reduce the time per iteration, i.e., a single kernel invocation, while the time to store/load to device memory in-between iterations remains the same. To summarize, PERKS reduces what amounts to be Amdahl's law serial portion of the solver, and hence the more optimized a code, i.e., a faster parallel portion, the higher the speedup that would be attributed to PERKS.

\textbf{PERKS in Distributed Computing:} PERKS in this paper is demonstrated on a single GPU. In distributed applications that require halo regions (e.g., stencils), PERKS can potentially be used on top of communication/computation overlapping schemes~\cite{7384347,9150372}. In overlapping schemes, the boundary points that are computed in a separate kernel would not be cached, while the kernel of the interior points would run as PERKS to cache the data of the interior points. PERKS could also be used with communication-avoiding algorithms (e.g., communication-avoiding Krylov methods~\cite{DBLP:conf/sc/IdomuraIAI20})

\textbf{Practicality of PERKS:} A wide range of iterative solvers (particularly iterative stencils) can be written as PERKS. However, it should be mentioned that there are applications at which the time stepping loop (on the host) is comprised of different GPU kernel. For instance, in production libraries, conjugate gradient (and Krylov solvers in general) are typically implemented as different kernels corresponding to different steps in the algorithm. In such case the kernels have to be fused~\cite{DBLP:conf/sc/WahibM14,DBLP:conf/ics/GysiGH15} first before transforming them to PERKS.

\textbf{Use of Registers:} PERKS uses registers and shared memory for caching data in-between time steps. It should be noted that there are no guarantees that the compiler releases all the registers after the compute portion in each iteration is finished (with Nvidia's nvcc compiler we did not observe such inefficiency). If such register reuse inefficiency exists, imperfect register reuse by the compiler could result in fewer registers being available for caching and leaves only shared memory to be used for caching. PERKS would not be effective if the target kernel consumes all on-chip resources (both register file and shared memory) even in its minimal occupancy.

\textbf{Iterative Solvers as PERKS:} While this paper's focus is to demonstrate PERKS model for iterative stencils and Krylov subspace methods (conjugate gradient), the discussion in this section (and paper in general) is applicable to a high degree for other types of iterative solvers. That is since PERKS is not much concerned with the implementation of the solver and only loads/stores the domain (or a subset of it) before/after the solver part in the kernel, under resource constraints. Iterative solvers that use the same flow expressed in Figure~\ref{fig:exemodel} can, in principle, be ported to PERKS (with relative ease). Generally speaking, the porting process is as follows: move the time step outside the kernel to be inside the kernel, add grid synchronization to ensure dependency, and store/load a portion of the input or output to cache: either shared memory and/or register (using register arrays). More details on porting kernels to PERKS are in Section~\ref{sec:enduser}.

\section{{Porting Solvers to PERKS}}\label{sec:imp}
Transforming the existing iterative solvers to PERKS is straightforward. This section first explains briefly how end-users can transform or port their iterative solvers to PERKS. Next, we elaborate on how we implemented memory-bound iterative methods (namely 2D/3D stencils and a conjugate gradient solver) as PERKS. 

\subsection{Transforming Kernels to PERKS: the End-user Perspective}\label{sec:enduser}
\begin{comment}
Transforming the existing iterative solvers to PERKS is fairly straightforward. This section first briefly discusses the process by which how the end-user could approach transforming kernels to PERKS. Next we discuss how we transformed the stencil and conjugate gradient kernels to PERKS to use in our experimentation. Finally, we briefly discuss CUDA-specific considerations when transforming kernels to PERKS.
\end{comment}

\subsubsection{Identifying the minimal concurrency of the kernel}\label{sec:userminocc}
The end-user {can rely on CUDA APIs~\footnote{cudaOccupancyMaxActiveBlocksPerMultiprocessor.}~\cite{nvidia2021api} to get the max concurrently running parameters. For even better performance, the end-user only} needs to reduce the device occupancy to \revision{its} minimum (while maintaining performance) via manual tuning of the kernel launch parameters or using auto-tuning tools~\cite{kerneltuner,shende2006tau,adhianto2010hpctoolkit}. 
\subsubsection{Porting a Kernel to become PERKS}
As Listing~\ref{Fig:codeperk22d5pt} shows, PERKS does not modify the computation; the manually written code to move the time loop inside the kernel and load/store to cache is straightforward. Alternatively, though outside this paper's scope, we point out the possibility of simplifying the process of converting a kernel to PERKS by using source-to-source translation, C++ templates, or domain specific languages.
\subsubsection{What to Cache}
The end-user can use a profiler, offline, to decide on what data arrays to cache by identifying the arrays that generate the most traffic to/from global memory. In many iterative solvers, profiling is not even needed since the algorithm clearly implies the main data array(s) causing the highest traffic (e.g., the matrix $A$ in conjugate gradient and the discretized domain in stencil applications).
\subsubsection{Where to Cache}
The end-user would simply use the unused shared memory for caching. For additional performance benefits, advanced users can choose to also cache in registers by manually identifying the adequate number of registers that can be used for caching, without causing register spilling (we provide a Python script to automate this process), or by following the trace of existing on-chip resources management research~\cite{vijaykumar2016zorua,li2015automatic}. We anticipate the possibility of automating this step by source-to-source translation or domain-specific languages so that this step of using on-chip resources could be as easy as adding a persisting range in the domain, similar, in principle, to the 
method of using L2 cache residency control in A100~\cite{nvidiaa100},\revision{ except that l2 cache residency control does not guarantee the data is definitely persistent~\cite{guide2020cuda}}.
%\subsubsection{Fine-level tuning}
%The existing steps provide a good speedup for most sceneries. Yet, smaller occupancy for caching a larger domain might be more profitable at specific domain sizes, especially at a smaller domain. An end-user can further conduct fine-level tuning to further reduce occupancy for better performance. 
%}

\subsection{{Transforming} Stencil Kernels to PERKS}
\subsubsection{{Stencil Kernel}}
\revision{We use SHM~\cite{maruyama2014optimizing} implementation as baseline. In SHM,} 3D stencil implementation uses the standard shared memory implementation where 2D planes (1D planes in 2D stencils) are loaded one after the other in shared memory. Each thread computes the cells in a vertical direction~\cite{8820786,10.1145/1513895.1513905}. In our PERKS implementation, before the compute starts, planes that already have the data cached from the previous time step do not load from global memory. 
\subsubsection{\revision{Porting the Stencil Kernel}}
We do not interfere with compute; only after the computation is finished do we store the results in the registers/shared memory. As Listing~\ref{Fig:codeperk22d5pt} shows, after adjusting to handle the input and output of the computation part of the kernel. \revision{We exchange halo region (inter thread block dependency data) between time steps.} To ensure coalesced memory accesses in the halo region, we transpose the vertical edges of the halo region. \revision{Also, we reuse the on-chip resources for caching as soon as the data is consumed. }Finally, \revision{since} the original kernel uses shared memory~\cite{8820786,10.1145/1513895.1513905} \revision{and} registers~\cite{chen2019versatile} to optimize stencils, we use the version of the output residing in shared memory or registers at the end of each time step as an already cached output. In this way, we avoid an unnecessary copy to shared memory and registers that we would use for caching. 
% \\
\begin{comment}
\begin{lstlisting}[language=C,rulecolor=\color{black}, basicstyle=\scriptsize, caption=2D 5-pt stencil implemented as PERKS,label={Fig:codeperk22d5pt},captionpos=t]
__global__ void 2d5pt_PERKS(ptr_in, ptr_out){...
for(k=0; k<timestep; k++){...
    switch (Source(ptr_in)){ 
    case FromSM:  load(sm_cache,sm_in);  break;
    case FromReg: load(reg_cache,sm_in);   break;
    default:      load(ptr_in,sm_in);}
    2d5pt_Compute(sm_in,reg_out);
    switch (Destination(ptr_out)){ 
    case ToSM:    store(reg_out,sm_cache); break;
    case ToReg:   store(reg_out,reg_cache);break;
    default:      store(reg_out,ptr_out);}
    ...
    //resolve dependency in halo region of TBs
    //(part of the baseline code; omitted for space)
    grid.sync();
    ...}
...}

__device__ void 2d5pt_Compute(sm_in, reg_out){ 
    x = threadIdx.x; 
    t[IPT+2]; //IPT: items per thread
    for(y=0; y< IPT+2; y++)
       t[y]=sm_in[x, y+ind_y-1]; 
    for( y=0; y< IPT; y++){
        reg_out[y]=sm_in[x+ind_x-1,y+1+ind_y]*WEST
                  +sm_in[x+ind_x+1,y+1+ind_y]*EAST 
                  +t[y-1+1]*SOUTH 
                  +t[y+1]*CENTER 
                  +t[y+1+1]*NORTH;
    }
}
\end{lstlisting}

\end{comment}

  % switch (Source(ptr_in)){ 
  % case FromSM:  <@\bf{load}@>(sm_cache,sm_in);     break;
  % case FromReg: <@\bf{load}@>(reg_cache,sm_in);    break;
  % default:      <@\bf{load}@>(ptr_in,sm_in);}
%%%%%%%%%%%%%%%%%%%%%%%%%%scriptsize
\lstset{
 	language = C++, breaklines = true, breakindent = 10pt, lineskip={-1pt}, basicstyle = \rmfamily\footnotesize, commentstyle = {\itshape \color[cmyk]{1,0.4,1,0}}, classoffset = 0, keywordstyle = {\bfseries \color[cmyk]{0,1,0,0}}, stringstyle = {\ttfamily \color[rgb]{0,0,1}}, frame = trbl, framesep=0pt, numbers = left, stepnumber = 1, xrightmargin=12pt, xleftmargin=0pt, numberstyle = \tiny, tabsize = 1, captionpos = t, directivestyle={\color{black}},  emph={int,char,double,float,unsigned, int3, float4, float2}, emphstyle={\color{blue}},
}
\lstset{escapeinside={<@}{@>}}

\begin{figure}[t]
\centering
\begin{minipage}[c]{0.5\textwidth}
% \footnotesize
\begin{lstlisting}[caption = {Iterative 2D 5-pt stencil implemented in PERKS.}, label = Fig:codeperk22d5pt]
__global__ void 2d5pt_PERKS(ptr_in, ptr_out)
{
  ...
  for(k=0; k<timestep; k++)
  {
    ...
    //Use branch to control the source 
    <@\bf{load}@>(bid,tid,ptr_in,
      [ptr_in|sm_cache|reg_cache]->sm_in);
    <@\bf\hllightgray{2d5pt\_Compute}@>(sm_in,reg_out);
    //Use branch to control the destination
    <@\bf{store}@>(bid,tid,ptr_out,
      reg_out->[ptr_out|sm_cache|reg_cache]);
    ...
    <@\bf{grid.sync}@>();
    ...
  }
  ...
}

__device__ void <@\bf\hllightgray{2d5pt\_Compute}@>(sm_in, reg_out){ 
  x = threadIdx.x; 
  t[IPT+2]; //IPT: items per thread
  for(y=0; y< IPT+2; y++)
    t[y]=sm_in[x, y+ind_y-1]; 
    for( y=0; y< IPT; y++){
      reg_out[y]=sm_in[x+ind_x-1,y+1+ind_y]*WEST
        +sm_in[x+ind_x+1,y+1+ind_y]*EAST 
        +t[y-1+1]*SOUTH 
        +t[y+1]*CENTER 
        +t[y+1+1]*NORTH;
    }
}
\end{lstlisting}
\end{minipage}
\end{figure}
\begin{comment}

\begin{figure}[t]
\centering
\begin{minipage}[c]{0.5\textwidth}
\begin{lstlisting}[caption = {2D 9-pt stencil implemented in PERKS.}, label = Fig:codeperk22d9pt]
__global__ void 2d9pt_PERKS(ptr_in, ptr_out){...
for(k=0; k<timestep; k++){...
  switch (Source(ptr_in)){ 
  case FromSM:  <@\bf{load}@>(sm_cache,sm_in);     break;
  case FromReg: <@\bf{load}@>(reg_cache,sm_in);    break;
  default:      <@\bf{load}@>(ptr_in,sm_in);}
  <@\bf\hllightgray{2d9pt\_Compute}@>(sm_in,reg_out);
  switch (Destination(ptr_out)){ 
  case ToSM:    <@\bf{store}@>(reg_out,sm_cache);  break;
  case ToReg:   <@\bf{store}@>(reg_out,reg_cache); break;
  default:      <@\bf{store}@>(reg_out,ptr_out);}
  ...
  //resolve dependency in halo region of TBs
  //part of the baseline code; omitted for space
  grid.sync();
  ...}
...}

__device__ void <@\bf\hllightgray{2d9pt\_Compute}@>(sm_in, reg_out){ 
    x = threadIdx.x; 
    t[IPT+4]; //IPT: items per thread
    for(y=0; y< IPT+4; y++)
        t[y]=sm_in[x, y+ind_y-2]; 
    for( y=0; y< IPT; y++){
      reg_out[y]=sm_in[x+ind_x-1,y+1+ind_y]*WES0
        +sm_in[x+ind_x-2,y+1+ind_y]*WEST1
        +sm_in[x+ind_x+1,y+1+ind_y]*EAST0 
        +sm_in[x+ind_x+2,y+1+ind_y]*EAST1 
        +t[y-1+2]*SOUTH0+t[y-2+2]*SOUTH1 
        +t[y+2]*CENTER 
        +t[y+1+2]*NORTH0+t[y+2+2]*NORTH1;
    }
}
\end{lstlisting}
\end{minipage}
\end{figure}
\end{comment}
%This is normal text.
%<@\bf\textcolor{red}{red text}@>
%More.
%%%%%%%%%%%%%%%%%%%%%%%%%
\subsection{{Transforming} the Conjugate Gradient Solver to PERKS}\label{sec:impCg}
\subsubsection{\revision{Conjugate Gradient Kernel}}
For simplicity and accessibility, we use the Conjugate Gradient (CG) solver implementation that is part of the CUDA SDK samples (\emph{conjugateGradientMultiBlockCG}~\cite{nvidia2021sample}). Since the implementation of SpMV in the CG sample is relatively naive, we use the highly optimized merge-based SpMV~\cite{merrill2016merge} that is part of the C++ CUB~\cite{nvidia2019cub}  library in the CUDA Toolkit~\cite{cudaToolkit}, as it fits naturally with the caching scheme in PERKS. We do not discuss the details of merge-based SpMV due to the space limit. The reader can refer to details in~\cite{merrill2016merge}. 

% \begin{comment}
\begin{comment}
\begin{lstlisting}[language=C,rulecolor=\color{black}, basicstyle=\scriptsize, caption=2D 5-pt stencil implemented as PERKS,label={Fig:codeperk22d5pt},captionpos=t]
__global__ void 2d5pt_PERKS(ptr_in, ptr_out){...
for(k=0; k<timestep; k++){...
    switch (Source(ptr_in)){ 
    case FromSM:  load(sm_cache,sm_in);  break;
    case FromReg: load(reg_cache,sm_in);   break;
    default:      load(ptr_in,sm_in);}
    2d5pt_Compute(sm_in,reg_out);
    switch (Destination(ptr_out)){ 
    case ToSM:    store(reg_out,sm_cache); break;
    case ToReg:   store(reg_out,reg_cache);break;
    default:      store(reg_out,ptr_out);}
    ...
    //resolve dependency in halo region of TBs
    //(part of the baseline code; omitted for space)
    grid.sync();
    ...}
...}

__device__ void 2d5pt_Compute(sm_in, reg_out){ 
    x = threadIdx.x; 
    t[IPT+2]; //IPT: items per thread
    for(y=0; y< IPT+2; y++)
       t[y]=sm_in[x, y+ind_y-1]; 
    for( y=0; y< IPT; y++){
        reg_out[y]=sm_in[x+ind_x-1,y+1+ind_y]*WEST
                  +sm_in[x+ind_x+1,y+1+ind_y]*EAST 
                  +t[y-1+1]*SOUTH 
                  +t[y+1]*CENTER 
                  +t[y+1+1]*NORTH;
    }
}
\end{lstlisting}

\end{comment}

  % switch (Source(ptr_in)){ 
  % case FromSM:  <@\bf{load}@>(sm_cache,sm_in);     break;
  % case FromReg: <@\bf{load}@>(reg_cache,sm_in);    break;
  % default:      <@\bf{load}@>(ptr_in,sm_in);}
%%%%%%%%%%%%%%%%%%%%%%%%%%scriptsize
\lstset{
 	language = C++, breaklines = true, breakindent = 10pt, lineskip={-1pt}, basicstyle = \rmfamily\footnotesize, commentstyle = {\itshape \color[cmyk]{1,0.4,1,0}}, classoffset = 0, keywordstyle = {\bfseries \color[cmyk]{0,1,0,0}}, stringstyle = {\ttfamily \color[rgb]{0,0,1}}, frame = trbl, framesep=0pt, numbers = left, stepnumber = 1, xrightmargin=12pt, xleftmargin=0pt, numberstyle = \tiny, tabsize = 1, captionpos = t, directivestyle={\color{black}},  emph={int,char,double,float,unsigned, int3, float4, float2}, emphstyle={\color{blue}},
}
\lstset{escapeinside={<@}{@>}}

\begin{figure}[t]
\centering
\begin{minipage}[c]{0.5\textwidth}
% \footnotesize
\begin{lstlisting}[caption = {Iterative Sparse matrix–vector multiplication with merge-based SpMV~\cite{merrill2016merge} implemented in PERKS.}, label = Fig:perkspmv]
__global__ void Iterative_SpMV_PERKS(...)
{
  ...
  while(...)
  {
    //<@{merge-based SpMV~\cite{merrill2016merge}}@> contains:
    //<@\bf{search}@>: determine the workload range
    //<@\bf\hllightgray{SpMV\_Compute}@>: performs SpMV within range
    ...
    //determine current thread block workload
    <@\bf{get}@>([cached_tb_range|<@\bf{search}@>(gm_matrix)]-> tb_range);
    //load tb_range of matrix to shared memory
    <@\bf{load}@>(tb_range,[sm_matrix|gm_matrix]->tb_sm_matrix);
    ...
    //determine current thread workload
    <@\bf{get}@>([cached_t_range|<@\bf{search}@>(tb_matrix)]->t_range);
    <@\bf\hllightgray{SpMV\_Compute}@>(tb_sm_matrix, t_range);
    ...
    <@\bf{grid.sync}@>();
    ...
  }
  ...
}
\end{lstlisting}
\end{minipage}
\end{figure}
\subsubsection{{Porting the Conjugate Gradient Kernel}}
We do not change the implementation or algorithm of the merge-based SpMV since PERKS does not necessitate changes in the underlying algorithm. For merge-based SpMV, we cache the matrix $A$ since it is the largest data array in the solver. To further improve performance, we also cache the residual vector {\footnotesize$r$} and the intermediate results. The merge-based SpMV~\cite{merrill2016merge} in CUB~\cite{nvidia2019cub} is composed of two steps: $search$ and $compute$. The search step is done twice. The search step first finds the workload for each thread block, and then finds the workload for each thread inside a thread block. The search result for the thread block workloads in global memory is saved since the matrix is static throughout the entire iteration. The second search (thread-level) is conducted in shared memory. Those two steps repeatedly generate intermediate data that we cache, in addition to the matrix $A$. Listing~\ref{Fig:perkspmv} shows code sample of PERKS based Iterative SpMV, which can be extended to a conjugate gradient solver.

Merge-based SpMV originally uses small thread blocks, i.e., $64$ threads per TB. This introduces a high volume of concurrently running thread blocks per streaming multiprocessor. To reduce the device occupancy while maintaining performance, we increased the TB size to $128$ and slightly changed the memory access order to accommodate the larger TB size.

\subsection{PERKS and CUDA Considerations}

\subsubsection{Restrictions of Synchronization APIs}
PERKS relies on cooperative groups APIs~\cite{nvidia2021api} (supported since CUDA 9.0). Currently, the APIs do not allow over-subscription, i.e., one needs to explicitly assign workload to blocks and threads to expose enough parallelism to the device. However, it is worth mentioning that this API does not limit the flexibility, as different kernels can still run concurrently in a single GPU, as long as they as a whole doesn't exceed the hardware limitation.
\subsubsection{New Features in Nvidia Ampere}
The Nvidia Ampere generation of GPUs introduced two new features that have the potential to improve the performance of PERKS. Namely, asynchronous copy for shared memory and L2 cache residency control~\cite{nvidiaa100}. When testing asynchronous copy to cache in PERKS, we did not observe noticeable performance difference. For L2 cache residency control, we experimented with setting the input and halo region to be persistent {in stencils}. We observed a $8\%$ slowdown and no change in performance, respectively. Accordingly, we do not include those new features in our PERKS implementations.
% \begin{comment}

% \begin{comment}  
\subsubsection{Register pressure in PERKS} 
One concern with PERKS is that kernels might run into register pressure if the compiler is not optimally reusing registers for different time steps, potentially affecting concurrency and penalizing performance. To illustrate this issue, take a high register-pressure 2D 25-point double precision Jacobi stencil as an example. The shared memory optimized baseline version (SHM) uses $78$ registers per thread, yet the PERKS version uses $112$ registers\footnote{We gathered the number of registers used by finding the maximum number of registers available as cache before spilling with "\_\_launch\_bounds\_\_" instruction. Register spilled can be indicated by '-Xptxas "-v -dlcm=cg"' flag.}. Similar behavior is also observed in other stencil benchmarks. Reducing the occupancy while maintaining the concurrency --as mentioned in the previous section-- reduces the impact of this compiler's inefficiency in register reuse in all the benchmarks we report in the results section. In the above example, at worst, $48$ registers among the maximum available $178$ registers per thread could not be used for caching data; it neither harms concurrency nor triggers register spilling. 
% \end{comment}

% \end{comment}
% \begin{table}[t]
%     \centering
%     \caption{Register information of single precision 2d5pt and double precision 2ds25pt}
%     \begin{tabular}{|c|c|c|}
%     \hline
%         TYPE & 2d5pt(f)  & 2ds25pt(d)  \\\hline
%         SM-OPT            & 32  & 78 \\
%         PERKS(SM-OPT)                    & 32  & 112\\
%         PERKS(SM-OPT) with $D_{cache}^{sm}$& 32  & 126\\\hline
%     \end{tabular}
%     \label{tab:registeranalysis}
% \end{table}

\section{Performance Analysis}\label{sec:simplepermodel}
%\lingqi{QUESTION: seems like peak performance not as important as concurrency analysis}

%\lingqi{Question: we need peak and conc}
%\chen{show the purpose of the performance model at the first sentences including why the perf. model is needed in this paper.}
%\begin{itemize}[leftmargin=3mm]
%\revision{\chen{In this section, we build a performance model to show the theoretical peak performance of PERK. Also, we can analyze the impact of different parameters on the performance of PERK. Hence, we can qualify the implementation quality by comparing our performance with the projected peak results.}

In this section, we propose a performance model that serves the following purposes. First, we propose a projection of achievable performance that we compare with measured results to detect abnormal behavior or implementation shortcomings. We relied on this projection in the analysis of our PERKS implementation quality (Section~\ref{sec:peakperform}).
%{\item\lingqi{Provide simple model to project performance of PERKS (Section~\ref{sec:projectionmodel} or Equation~\ref{eqt:projectionLarge})} \wahib{@Lingqi@:this point is similar to the previous point, so please merge.}}
%\item Identifying limitations of caching schemes and shortcomings of caching policies (Section~QQ).\lingqi{I don't know what to do with it}
Second, we identify the bounds on reducing concurrency before performance regression and use concurrency to explain potential optimizations for further performance improvement (Section~\ref{sec:conan}). \revision{It is worth mentioning that the concurrency analysis is not a requirement for porting kernels to PERKS; we use the analysis to understand the feasibility of PERKS in practice, and address its implication on performance.}
\subsection{Overview}
%\wahib{Note to self: remove parameter $M$ and adjust accordingly.}
This performance model relies on three performance attributes: a) measured performance $\mathbb{M}$ of our PERKS implementation, b) the projected peak performance $\mathbb{P}$ achievable on a given GPU,  and c) the efficiency function $\mathfrak{E()}$ describing the efficiency of the given kernel running on the device.
%\revision{gap} between $\mathbb{P}$ and $\mathbb{M}$. 
More specifically, $\mathfrak{E()}$ is a function of the concurrency exposed by the software $\mathbb{C}_{sw}$ and the concurrency required by the hardware $\mathbb{C}_{hw}$. The relation of measured performance to projected peak performance becomes:
%Eff(C_{code},C_{hw})  
%\color{blue}
\begin{equation}\footnotesize
   \mathbb{M}  =  \mathbb{P}\times \mathfrak{E(} \revision{\mathbb{C}_{sw}},\mathbb{C}_{hw}\mathfrak{)}
    \label{eqt:basic}
\end{equation}
%Section~\ref{sec:peakperform}
We discuss projected peak performance in the following section. A detailed discussion of the efficiency and concurrency functions is in Section~\ref{sec:perconcurrency}.

\subsection{Projecting Peak Achievable Performance}\label{sec:peakperform}
We rely on the figure of merit as the performance metric in this analysis. In stencils, we use the giga-cells updated per second (GCells/s)~\cite{DBLP:conf/cgo/MatsumuraZWEM20,chen2019versatile}. Given the memory-bound nature of the conjugate gradient solver, we directly use sustained memory bandwidth as a metric, following other works on conjugate gradient~\cite{anzt2020ginkgo}. Due to space limitations, this section mainly focuses on stencils to explain the performance analysis. Without loss of generality, the analysis is applicable to other cases (ex: conjugate gradient) by adjusting the performance metric and code concurrency accordingly.

We use a simple performance model inspired by the roofline model~\cite{ofenbeck2014applying,kim2011performance}. The model's utility is to project the upper bound on performance based on the reduction of global memory traffic. This model, in turn, helps us in this paper to identify performance gaps in our PERKS implementation and later inspect the reasons for those gaps.% (e.g., register spilling). 

% \subsubsection{Baseline}
% To show that the baseline implementation is enough optimized, the peak performance for baseline is analyzed in this subsection. By assuming that the bottleneck is global memory bandwidth, and all memory transactions are coalesced, baseline performance can be computed based on the memory bandwidth. 
In a kernel implemented as PERKS, the bottleneck could either be the global memory bandwidth or the shared memory bandwidth (if the PERKS caching scheme moves the bottleneck to become the shared memory bandwidth). We don't assume the registers to be a bottleneck since we assume that as long as we ensure that no register spilling occurs, we avoid register pressure.

We assume a total domain of size $\mathbb{D}$ bytes, the cached portion to be $\mathbb{D}_{cache}$ bytes, and the uncached portion to be {\footnotesize$\mathbb{D}_{uncache}=\mathbb{D}-\mathbb{D}_{cache}$} bytes. The cached portion of the domain data would be divided between registers and shared memory (since we cache in both registers and shared memory): {\footnotesize$\mathbb{D}_{cache}=\mathbb{D}^{sm}_{cache}+\mathbb{D}^{reg}_{cache}$}. For $N$ time steps, \revision{a}ssuming the number of bytes stored to global memory in each time step is $\mathbb{S}_{gm}$ and the number of bytes loaded is $\mathbb{L}_{gm}$, the total global memory bytes accessed $\mathbb{A}_{gm}$ becomes:
%\begin{equation}\footnotesize
%    \mathbb{L}_{global}(i)= 
%\begin{cases}
%    \mathbb{D}_{total},& \text{if } i= 0\\
%    \mathbb{D}_{global}, & \text{otherwise}
%\end{cases}
%\end{equation}
%\begin{equation}\footnotesize
%    \mathbb{S}_{global}(i)= 
%\begin{cases}
%    \mathbb{D}_{total},& \text{if } i= n-1\\
%    \mathbb{D}_{global}, & \text{otherwise}
%\end{cases}
%\end{equation}
\begin{equation}\footnotesize
\begin{aligned}
    %\mathbb{A}^{gm}(\mathbb{D})&=\sum_{i=0}^{n-1}(\mathbb{L}^{gm}(i)+\mathbb{S}^{gm}(i))\\
    %&=2\cdot n\cdot \mathbb{D}_{uncache} + 2\cdot  \mathbb{D}_{cache}
    %\mathbb{A}_{gm}(\mathbb{D})&=N\cdot(\mathbb{L}_{gm}+\mathbb{S}_{gm})\\
    %&=2\cdot N\cdot \mathbb{D}_{uncache} + 2\cdot  \mathbb{D}_{cache}
    \mathbb{A}_{gm}(\mathbb{D})=N\cdot(\mathbb{L}_{gm}+\mathbb{S}_{gm})
    &=2\cdot N\cdot \mathbb{D}_{uncache} + 2\cdot  \mathbb{D}_{cache}    
%    \\
%    &\stackrel{\text{n is large enough}}{\approx} 2\cdot n\cdot \mathbb{D}_{global}
\end{aligned}
\end{equation}

When the kernel is bounded by global memory bandwidth, i.e., the volume of cached data does not move the bottleneck from global memory to shared memory, for the global memory bandwidth of $\mathbb{B}_{gm}$ and data type size of $\mathfrak{S}(type)$, the time $\mathbb{T}_{gm}(\mathbb{D})$ for accessing the global memory becomes:%can be computed base on Equation~\ref{eqt:time}:

\begin{equation}\footnotesize
    \begin{aligned}
        %\mathbb{T}_{gm}(\mathbb{D})=\frac{{\mathbb{A}_{gm}(\mathbb{D})\cdot \mathfrak{S}(type)}}{{\mathbb{B}_{gm}}}
                \mathbb{T}_{gm}(\mathbb{D})={{\mathbb{A}_{gm}(\mathbb{D})\cdot \mathfrak{S}(type)}}/{{\mathbb{B}_{gm}}}
    \end{aligned}
    \label{eqt:time_gm}
\end{equation}

In the case when the kernel is bounded by shared memory bandwidth, i.e., the volume of data cached in shared memory moves the bottleneck to be the shared memory bandwidth, the total shared memory (in bytes) accessed $\mathbb{A}_{sm}$ becomes: %we assume the algorithm that we plan to apply PERKS be $\mathbb{ALG}$. PERKS won't influence the shared memory access amount of $\mathbb{ALG}$ per time step, which can be assumed as $\mathbb{A}^{sm}(\mathbb{ALG})$. 
%But some additional memory access within cached with shared memory region $\mathbb{A}^{sm}(\mathbb{D}^{sm}_{cache})$ exists if caching data in shared memory. Similar to global memory access we can get: 

\begin{comment}
\begin{equation}\footnotesize
    \mathbb{L}_i(sm)= 
\begin{cases}
    0,& \text{if } i= 0\\
    \mathbb{D}_{sm}, & \text{otherwise}
\end{cases}image.png
\end{equation}
\begin{equation}\footnotesize
    \mathbb{S}_i(sm)= 
\begin{cases}
    0,& \text{if } i= n-1\\
    \mathbb{D}_{sm}, & \text{otherwise}
\end{cases}
\end{equation}
\end{comment}

\begin{equation}\footnotesize
\begin{aligned}
    %\mathbb{A}_{sm}(\mathbb{D}_{cache}^{sm})&=N\cdot(\mathbb{L}_{sm}+\mathbb{S}_{sm})\\
    %				&=2\cdot (N-1)\cdot \mathbb{D}_{cache}^{sm}
    \mathbb{A}_{sm}(\mathbb{D}_{cache}^{sm})=N\cdot(\mathbb{L}_{sm}+\mathbb{S}_{sm})
    				=2\cdot (N-1)\cdot \mathbb{D}_{cache}^{sm}    				
    %\mathbb{A}^{sm}(\mathbb{D}_{cache})&=\sum_{i=0}^{n-1}(\mathbb{L}^{sm}(i)+\mathbb{S}^{sm}(i))\\
    %				&=2\cdot (n-1)\cdot \mathbb{D}^{sm}_{cache}
\end{aligned}
\end{equation}

%Additionally, given that the global memory bandwidth is $\mathbb{B}^{sm}$ and the data type size is $\mathfrak{S}(type)$, the time $\mathbb{T}$ for accessing the global memory can be computed base on Equation~\ref{eqt:time}:
Assuming $\mathbb{A}_{sm}(\mathbb{KERNEL})$ to be the shared memory originally used by the kernel, e.g., shared memory used in the baseline implementation of a stencil kernel to improve the locality, and $\mathbb{B}_{sm}$ to be the shared memory bandwidth, the time $\mathbb{T}_{sm}(\mathbb{D})$ for accessing the shared memory becomes:

\begin{equation}\footnotesize
    \begin{aligned}
        %\mathbb{T}_{sm}(\mathbb{D}_{cache}^{sm})=\frac{{(\mathbb{A}_{sm}(\mathbb{D}_{cache}^{sm})+\mathbb{A}_{sm}(\mathbb{KERNEL}))\cdot \mathfrak{S}(type)}}{{\mathbb{B}_{sm}}}
        \mathbb{T}_{sm}(\mathbb{D}_{cache}^{sm})=\{{{(\mathbb{A}_{sm}(\mathbb{D}_{cache}^{sm})+\mathbb{A}_{sm}(\mathbb{KERNEL}))\cdot \mathfrak{S}(type)}}\}/{{\mathbb{B}_{sm}}}
    \end{aligned}
    \label{eqt:time_sm}
\end{equation}

%The time $\mathbb{T}_{domain}$ for accessing the domain, Assuming perfect overlapping, in PERKS is:
%
%\begin{equation}\footnotesize
%    \begin{aligned}
%     \mathbb{T}_{domain}=\max(\mathbb{T}_{global},\mathbb{T}_{{sm}}
%    \end{aligned}
%    \label{eqt:maxlat}
%\end{equation}

%revert% 
%Next, we consider the unavoidable global memory accesses necessary to resolve the neighborhood dependencies in $\mathbb{D}_{cache}$, e.g., the halo region of stencils. In the cached portion of the domain, the halo region requires global memory accesses to resolve the dependencies. Assuming the global memory accesses, in bytes, for the halo region of the data computed by the boundary threads in thread blocks that are in the cached portion of the domain to be $\mathbb{A}_{gm}(\mathfrak{H}(\mathbb{D}_{cache}))$, the time for accessing those halo region $\mathbb{T}_{gm}(\mathfrak{H}(\mathbb{D}_{cache}))$ would be:

%\begin{equation}\footnotesize
%    \begin{aligned}
%          \mathbb{T}_{gm}(\mathfrak{H}(\mathbb{D}_{cache}))={\mathbb{A}(\mathfrak{H}(\mathbb{D}_{cache}))\cdot \mathfrak{S}(type)}/{\mathbb{B}_{gm}}
%    \end{aligned}
%    \label{eqt:perkhalo}
%\end{equation}
%revert ends% 
The projected best-case total time required for the PERKS kernel may be written as:

\begin{equation}\footnotesize
    \begin{aligned}
     \mathbb{T}_{PERKS}=\max(\mathbb{T}_{gm}(\mathbb{D}),\mathbb{T}_{sm}(\mathbb{D}_{cache}^{sm}))
    \end{aligned}
    \label{eqt:maxlat}
\end{equation}

Accordingly, the projected peak performance ($\mathbb{P}$ in Equation~\ref{eqt:basic}) for the $N$ time steps can be expressed as:

\begin{equation}\footnotesize
    \begin{aligned}
     %\mathbb{P}=\frac{\mathbb{D} \cdot N}{\mathbb{T}_{PERKS}}
     \mathbb{P}={\mathbb{D} \cdot N}/{\mathbb{T}_{PERKS}}
     %\mathbb{P}_{PERKS}=\frac{\mathbb{D} \cdot n}{\mathbb{T}_{PERKS}}
    \end{aligned}
    \label{eqt:maxpeak}
\end{equation}

%To further illustrate how this performance analysis works, we give two examples while computing $N=1000$ time-steps of a single precision 2D 5-point Jacobi stencil on A100. 

%To further illustrate how this performance analysis works, 
We give an example of computing $N=1000$ time-steps of a single precision 2D 5-point Jacobi stencil on A100. We use the domain size $\mathbb{D}=3072^2$; the total cache-able region is $\mathbb{D}_{cache}=3072\cdot2448$ leading to $\mathbb{T}_{gm}(\mathbb{D})=9900.70$ us. The total number of bytes for the halo accesses is $\mathbb{A}(\mathfrak{H}(\mathbb{D}_{cache}))=1000\cdot2\cdot 216\cdot(136\cdot2+256\cdot2)$. Thus 
% $\mathbb{T}^{gm}(\mathfrak{H}(\mathbb{D}_{cache}))=870.35$ us. 
 $\mathbb{T}^{gm}(\mathfrak{H}(\mathbb{D}_{cache}))= 871.22$ us. 
% So $\mathbb{P}_{PERKS}=3072^2\cdot 1000/\mathbb{T}_{PERKS}=876.16$ GCells/s, 
So $\mathbb{P}_{PERKS}=3072^2\cdot 1000/\mathbb{T}_{PERKS}= 876.09$ GCells/s.
\subsection{Concurrency and Micro-benchmarks}\label{sec:perconcurrency}
% \lingqi{note to self: stop here}
%\lingqi{note to self: rethink how to describe concurrency per thread}
%\lingqi{note to self: What is the concurrency used in this paper}
Reducing device occupancy increases the availability of resources to be used for caching in PERKS (as illustrated earlier in Figure~\ref{Fig:showcase}). On the contrary, reducing occupancy can lead to lower device utilization. To effectively implement PERKS, one has to reduce the occupancy as much as possible without scarifying performance. Inspired by the findings of Volkov~\cite{volkov2010better}, we assume that the efficiency function $\mathfrak{E}$ reaches its peak point when the code provides enough concurrency to saturate the device (irrespective of the occupancy):
\begin{equation}\footnotesize
    \mathfrak{E}(\mathbb{C}_{sw},\mathbb{C}_{hw})=100\%, \text{if } \forall \mathbb{C}_{sw}\geq \mathbb{C}_{hw}
\label{eqt:littlelaw}
\end{equation}

%The efficiency function $\mathfrak{E}()$ in this paper is used heuristically to ensure that PERKS won't harm performance. General cases when $\exists \mathbb{C}_{sw} < \mathbb{C}_{hw}$ is beyond the topic of this paper. 

Where $\mathbb{C}_{sw}(\mathbb{OP})$ is the minimum number of concurrently executable instructions of the operation $\mathbb{OP}$ exposed by the launched kernel, and $\mathbb{C}_{hw}(\mathbb{OP})$ is the maximum numbers of instructions of the operation $\mathbb{OP}$ that the device is capable of handling concurrently. Because this paper mainly focuses on memory bound applications, the $\mathbb{OP}$ referred to in this paper are limited to data access operations, i.e. global memory load/store $\mathbb{C}(\mathbb{GM})$, shared memory load/store $\mathbb{C}(\mathbb{SM})$, and L2 cache load/store $\mathbb{C}(\mathbb{L}2)$. 

%Specifically, $\mathbb{C}_{sw}(\mathbb{OP})$ denotes the max unblocking operations $\mathbb{OP}$ in software $sw$; $\mathbb{C}_{hw}(\mathbb{OP})$ denotes the minimal the operations $\mathbb{OP}$ that necessary to saturate a given device $hw$. Device level concurrency usually really large, we use stream multiprocessor level concurrency $\mathbb{C}_{sw}^{SM}$ and $\mathbb{C}_{hw}^{SM}$ for most of followning discussion. Because this paper mainly focus on memory bound applications, the $\mathbb{OP}$ referred in this paper is limited to memory access related, i.e. global memory access $\mathbb{C}(\mathbb{GM})$, shared memory access $\mathbb{C}(\mathbb{SM})$, and L2 cache access $\mathbb{C}(\mathbb{L}2)$. 

%When $\exists\mathbb{C}_{code}(\mathbb{OPT})<\mathbb{C}_{code}(\mathbb{OPT})$, the $\mathbb{OPT}$ could not saturate the device, the $\mathbb{M}$ would be far away from $\mathbb{P}$
\subsubsection{Measuring $\mathbb{C}_{sw}^{SM}$} $\mathbb{C}_{sw}^{SM}(\mathbb{OP})$, the kernel concurrency at the Streaming Multi-processor (SM) level, can be computed based on the concurrency exposed by the threads of a thread block $\mathbb{C}_{sw}^{TB}(\mathbb{OP})$ and number of concurrently running thread blocks per SM {\footnotesize$TB/SM$}:  {\footnotesize$\mathbb{C}_{sw}^{SM}(\mathbb{OP})=\mathbb{C}_{sw}^{TB}(\mathbb{OP})\cdot TB/SM$}.

%As an example, we use the code example in Fig~\ref{fig:ccrcyexmaple} to deduce the code concurrency of global memory load and stores $\mathbb{C}^{SM}_{sw}(\mathbb{GM})$ in a single Stream Multiprocessor to be: $\mathbb{C}^{SM}_{sw}(\mathbb{GM})=\mathbb{C}^{TB}_{sw}(\mathbb{GM})=256\cdot 2 = 2048$ load and store operations for type float. Note that the number of operations in this example is then adjusted to accounts for the effect of coalesced access on the actual number of operations (or transactions).

%  \begin{comment}
%  \begin{figure}
% \centering
% \begin{lstlisting}[language=C,gobble=2]
%  blocks = 80; thread/block = 256; block/SM = 1;
%  __global__ void Kernel(float*A,float*B,...)
%  {...
%    A[(256*bid)+tid] = B[(256*bid)+tid] * tid;
%  }
% \end{lstlisting}
% \caption{\label{fig:ccrcyexmaple} Example for concurrency counting in a CUDA kernel.}
% \end{figure}
% \end{comment}
 
\subsubsection{Measuring $\mathbb{C}_{hw}$} According to Little's Law~\cite{Little1961APF}, the hardware concurrency $\mathbb{C}_{hw}$ can be determined by the throughput $\mathbb{THR}$ and latency $\mathbb{L}$\cite{volkov2010better}: 
\begin{equation}\footnotesize
    \mathbb{C}_{hw}(\mathbb{OP})=\mathbb{THR(OP)} \cdot \mathbb{L(OP)}
\end{equation}

The throughput $\mathbb{THR}$ for data access operations are available in CUDA documentation~\cite{nvidia2019programming,amperewhite}. We measure the latency $\mathbb{L}$ with commonly used microbenchmarks~\cite{wong2010demystifying,mei2014benchmarking,zhang2020study}.% (The latency measurements are collected in the AD/AE appendix). %Throughput and latency of relevant operations on A100 and V100 are summarized in Table~\ref{tab:concurrency}. 

\subsection{Concurrency Analysis}
\label{sec:conan}
\begin{table}[t]
    \centering
    \caption{Concurrency analysis of global memory accesses of a single precision 2D-5point Jacobi stencil kernel running on A100 (1000 time-steps on $3072^2$ domain)}
    \resizebox{\linewidth}{!}
    {
    % \small
        \begin{tabular}{?c?c|c|c|c|c?}
        \tbhline
            \multirow{2}{*}{$\tfrac{TB}{SM}$}  & Used Reg. & Unused Reg. & GM Load & GM Store & Measured\\
              & /SM  &/SM    & op/SM & op/SM  & GCells/s\\\tbhline
            \textit{1} & 32KB  & 224KB  & 2580 & 2048 & 94.75\\
            \textit{2} & 64KB   & 192KB & 5160 & 4096 & 133.24\\
            \textit{8} & 256KB  & 0KB   & 20640 & 16384 & 138.29\\\tbhline
        \end{tabular}
    }
    \label{tab:concurrencyanalysis}
\end{table}
% \begin{table}[t]
%     \centering\small
%     \caption{Stencil benchmarks. A detailed description of the stencil benchmarks can be found in~\cite{zhao2019exploiting,rawat2016effective}}
%     \resizebox{\linewidth}{!}{
%     \begin{tabular}{|cccc|}
%     \hline
%         \multicolumn{4}{|c|}{ \textbf{Benchmark(Stencil Order, FLOPs/Cell)}}\\\hline
%         2d5pt(1,10) & 2ds9pt(2,18) & 2d13pt(3,26) & 2d17pt(4,34)\\
%         2d9pt(1,18) & 2d25pt(2,50)&3d7pt(1,14) & 3d13pt(2,26)\\
%         3d17pt(1,34) & 3d27pt(1,54)&poisson(1,38)&-\\ \hline
%     \end{tabular}}
%     \label{tab:stencilbenchmark}
% \end{table}
\begin{table*}[t]
    \centering
    \caption{Stencil benchmarks and domain sizes we use. A detailed description of the stencil benchmarks can be found in~\cite{zhao2019exploiting,rawat2016effective}. For fairness, the domain sizes we use are the minimum domain sizes that would saturate the device for different stencil benchmarks. The minimum domain sizes are identified by running each benchmark at different domain sizes and using the domain size after which the performance (in GCells/s) seize to improve}
    % \resizebox{\columnwidth}{!}
    {%
    \small
    \begin{tabular}{|c|cc|}\hline
        \textbf{Type}    &\multicolumn{2}{c|}{\textbf{2d stencils}}\\
         {\footnotesize(Stencil Order, FLOPs/Cell)} & \textbf{A100}             & \textbf{V100}           \\\hline 
        \textbf{2d5pt   (1,10)}    & $2304\times2304$          & $2048\times1280$        \\
        \textbf{2ds9pt  (2,18)}        & $2304\times2304$          & $2048\times1280$        \\
        \textbf{2d13pt  (3,26) }      & $4608\times3072$          & $2048\times2048$        \\
        \textbf{2d17pt  (4,34)}      & $3072\times2304$          & $4096\times2560$        \\
        \textbf{2d9pt  (1,18)}        & $2304\times2304$          & $2048\times1280$        \\
        \textbf{2d25pt (2,50)}       & $4608\times3072$          & $2048\times1280$        \\\hline
    \end{tabular}
    \begin{tabular}{|c|cc|}\hline
        \textbf{Type}    &\multicolumn{2}{c|}{\textbf{3d stencils}}\\
         {\footnotesize(Stencil Order, FLOPs/Cell)} & \textbf{A100}             & \textbf{V100}           \\\hline 
        \textbf{3d7pt (1,14)}    & $256\times288\times256$   & $128\times128\times128$ \\
        \textbf{3d13pt (2,26)}   & $256\times288\times256$   & $256\times320\times256$ \\
        \textbf{3d17pt (1,34)}   & $256\times288\times256$   & $160\times160\times256$ \\
        \textbf{3d27pt (1,54)}   & $256\times288\times256$   & $160\times160\times256$ \\
        \textbf{poisson (1,38)}   & $256\times288\times256$   & $160\times160\times256$ \\
         --- & ---  & ---\\
        \hline
    \end{tabular}
    }
    \label{tab:domain}
\end{table*}

 \begin{table*}[ht]
     \caption{Datasets for the Conjugate Gradient (CG) solver (from SuiteSparse~\cite{davis2011university})}
% \resizebox{\textwidth}{!}
{%
\small
     \begin{tabular}[t]{|cccc|cccc|cccc|}
     \hline
         \textbf{Code} & \textbf{Name~\cite{davis2011university}} &\textbf{Rows}& \textbf{NNZ} \\ \hline
         \textbf{D1}&Trefethen\_2000 & 2,000   & 41,906      \\
         \textbf{D2}&msc01440       & 1,440   & 46,270      \\
         \textbf{D3}&fv1            & 9,604   & 85,264      \\
         \textbf{D4}&msc04515       & 4,515   & 97,707      \\
         \textbf{D5}&Muu            & 7,102   & 170,134     \\
         \textbf{D6}&crystm02       & 13,965  & 322,905     \\
         \textbf{D7}&shallow\_water2 & 81,920  & 327,680     \\
 \hline
     \end{tabular}
     \begin{tabular}[t]{|cccc|cccc|cccc|}
     \hline
         \textbf{Code} & \textbf{Name~\cite{davis2011university}} &\textbf{Rows}& \textbf{NNZ} \\ \hline
         \textbf{D8}&finan512       & 74,752  & 596,992     \\
         \textbf{D9}&cbuckle        & 13,681  & 676,515     \\
         \textbf{D10}&G2\_circuit     & 150,102 & 726,674     \\
         \textbf{D11}&thermomech\_dM  & 204,316 & 1,423,116    \\
         \textbf{D12}&ecology2       & 999,999 & 4,995,991   \\
         \textbf{D13}&tmt\_sym        & 726,713 & 5,080,961   \\
         \textbf{D14}&consph         & 83,334  & 6,010,480   \\
\hline
     \end{tabular}
     \begin{tabular}[t]{|cccc|cccc|cccc|}
     \hline
         \textbf{Code} & \textbf{Name~\cite{davis2011university}} &\textbf{Rows}& \textbf{NNZ} \\ \hline
         \textbf{D15}&crankseg\_1     & 52,804  & 10,614,210  \\
         \textbf{D16}&bmwcra\_1       & 148,770 & 10,644,002  \\
         \textbf{D17}&hood           & 220,542 & 10,768,436  \\
         \textbf{D18}&BenElechi1     & 245,874 & 13,150,496  \\
         \textbf{D19}&crankseg\_2     & 63,838  & 14,148,858  \\
         \textbf{D20}&af\_1\_k101      & 503,625 & 17,550,675  \\ 
         \textbf{---}&---     & --- & --- \\ \hline
     \end{tabular}
     }
    
     \label{tab:matrxset}
 \end{table*}
%%The main benefit of performance model is that it could guide optimizations when un-reasonable slowdown is observed. In PERKS performance model is also necessary to ensure that applying PERKS don't harm performance. 

%Even though in this research we haven't encountered, we anticipate 2 scenarios that PERKS might harm performance: 1). The original kernel $\mathbb{KERNEL}$ has high register pressure that any modification might introduce register spill or already suffer from register spill issue, or 2). The $\mathbb{KERNEL}$ is under-optimized that need to rely on large amount of concurrently running threads to ensure performance. 

%Solving first issue is not the purpose of this research, readers can refer to existing register optimizations~\cite{Rawat:2018:AIR:3291656.3291718,Rawat:2018:ROS:3200691.3178500}; yet the second issue can be resolve by using concurrency analysis to guide code optimization. 

% \subsubsection{Use concurrency analysis to reduce thread blocks per stream multiprocessor $TB/SM$ to its minimum}
In this section, we briefly describe how we analyze the concurrency to reduce the occupancy of the original kernel in order to release resources for caching while sustaining performance. We conduct a static analysis to extract the data movement operations in the kernel. Note that we account for any barriers in the original kernels that could impact the concurrency of operations, i.e., we do not combine operators/instructions from before and after the barrier when we count the operators. Finally, we apply a simple model (Equation~\ref{eqt:littlelaw}) to identify the least occupancy we could drop to before the concurrency starts to drop. The results summarized in Table~\ref{tab:concurrencyanalysis} show that for a 2D 5-point Jacobi stencil kernel kernel, we could reduce the original occupancy to ${1/4}^{th}$ while maintaining performance.
%For example, we did a static analysis of a single precision 2D 5-point Jacobi stencil kernel to conduct a concurrency analysis. It is worth mentioning that the baseline kernel we use is fairly complex since it is highly optimized by using shared memory to reduce traffic to global memory to its minimum~\cite{maruyama2014optimizing}. 

% $\mathbb{C}_{sw=2d5pt,TB/SM=1}^{SM}(\mathbb{GM})=18512$B$>\mathbb{C}_{hw=A100}^{SM}(\mathbb{GM})$. Also $\mathbb{C}_{sw=2d5pt,TB/SM=1}^{SM}(\mathbb{SM})>10320$B$>\mathbb{C}_{hw=A100}^{SM}(\mathbb{SM})$

% \footnote{In this implementation, loading from global memory to shared memory happens before the computation starts.}
To understand the gap between the performances at 1 vs. 8 TB/SM  ($94.75/138.29=68.52\%$), we inspect the efficiency function $\mathfrak{E()}$. The number of concurrent global memory accesses and shared memory accesses in the 2D 5-point Jacobi stencil kernel are enough to saturate A100 when TB/SM=1. Accordingly, we get $\mathfrak{E}_(\mathbb{C}_{sw=j2d5pt,TB/SM\geq1},\mathbb{C}_{hw=A100})=1$, which would indicate that the observed gap in performance is not due to a drop in concurrency we did not model. While this confirms the effectiveness of the concurrency analysis (i.e., since the concurrency analysis resonates with the empirical measurements in Table~\ref{tab:concurrencyanalysis}), it does not uncover the source of the performance gap. Investigative profiling revealed that the concurrency for accesses in L2 cache, not global memory, is impacted by reducing occupancy on A100 in specific to the level that affects performance notably. More particularly, access to global memory for the halo region garners a high L2 cache hit rate. This effectively means that higher concurrency is necessary to saturate the L2 cache when hit rates are high. To confirm, we manually doubled the concurrency $\mathbb{C}^{TB}_{sw}$: the performance increased to 123.94 GCells/s with TB/SM=1 (from $68.52\%$ up to $89.6\%$). 
\color{black}

\begin{figure*}[t]
\centering
\includegraphics[width=\textwidth]{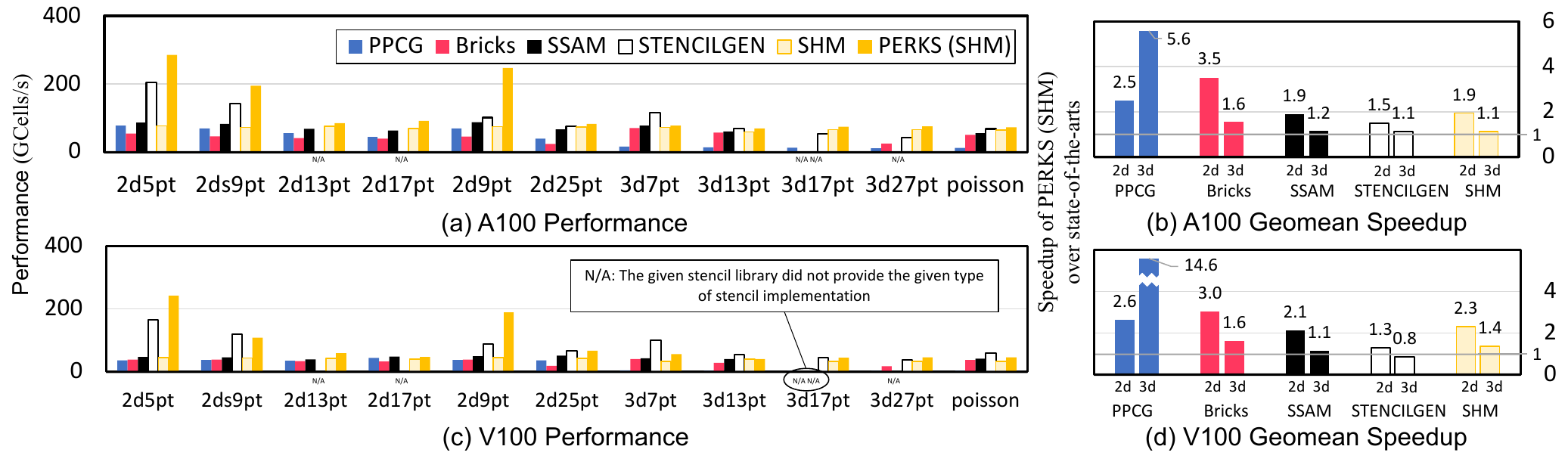}
\caption{\label{fig:perksresult}Comparison of PERKS(SHM~\cite{maruyama2014optimizing}) over a wide range of stencil libraries in a wide range of 2D/3D stencil benchmarks on A100 (top) and V100 (bottom) GPUs. (a) \& (c) in the left report the performance; (b) \& (d) in the right report the geometric mean speedup of PERKS(SHM) over other state-of-the-art implementations.
}
\end{figure*}
\begin{figure*}[t]
\centering
\includegraphics[width=\textwidth]{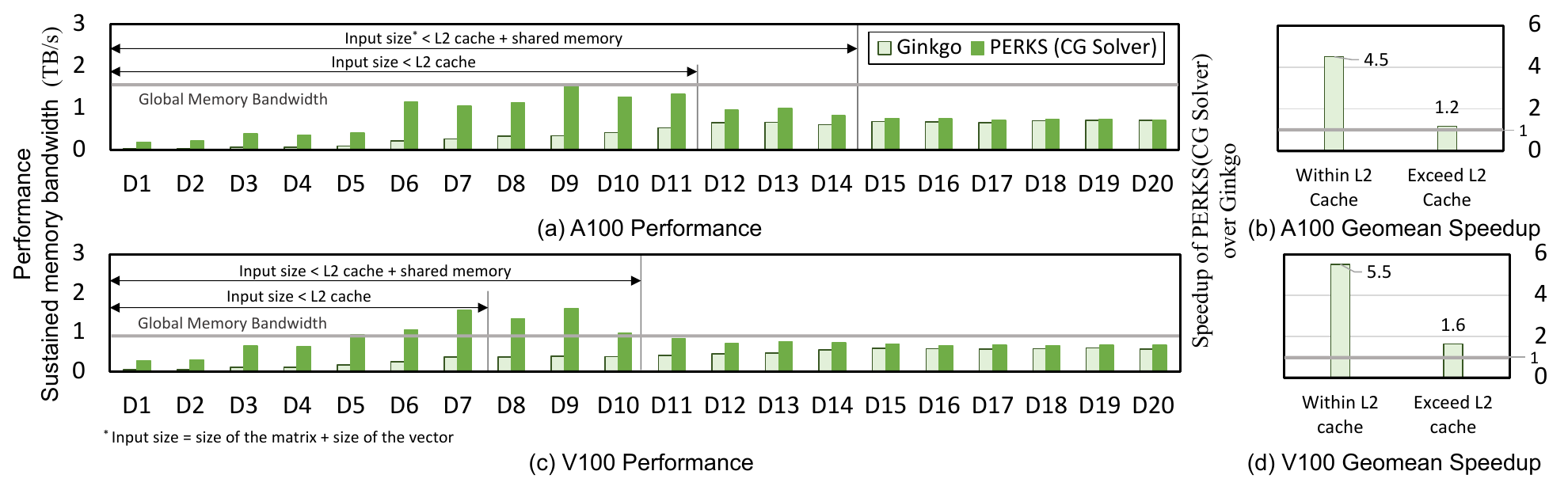}
\caption{\revision{Comparison of PERKS(CG Solver, CUDA Sample~\cite{nvidia2021sample}+CUB~\cite{merrill2016merge}) over Ginkgo~\cite{anzt2020ginkgo} on A100 (top) and V100 (bottom) GPUs in datasets (D1 to D20) listed in Table~~\ref{tab:matrxset}. (a) \& (c) in the left report the performance (sustained memory bandwidth); (b) \& (d) in the right report the geometric mean speedup of PERKS(CG Solver) over Ginkgo.% We observe a significant difference performance feature of PERKS(CG Solver) when the domain size is within the L2 cache and exceeds the L2 cache, so we report the geometric mean speedup separately.
}
}
\label{Fig:cg_compare}
\end{figure*}

\section{Evaluation}\label{sec:eval}
\subsection{Hardware and Software Setup}

The experimental results presented here are evaluated on the two latest generations of Nvidia GPUs: Volta V100 and Ampere A100 with CUDA 11.5 and driver version 495.29.05.

We run each evaluation ten times for all iterative stencils and conjugate gradient experiments. All experimental results reported are done in double precision.

\subsection{Benchmarks and Datasets}

\subsubsection{{Stencil Benchmarks}} 

To evaluate the performance of PERKS stencils, we perform a wide set of experiments on various 2D/3D stencil benchmarks (listed in Table~\ref{tab:domain}).

{We compare PERKS (w/ SHM~\cite{maruyama2014optimizing} as base) with a wide range of state-of-the-art stencil implementations/libraries: PPCG~\cite{verdoolaege2013polyhedral}, Bricks~\cite{zhao2019exploiting}, SSAM~\cite{chen2019versatile}, STENCILGEN~\cite{rawat2018domain}} and SHM~\cite{maruyama2014optimizing}. The implementations/libraries represent different classes of stencil optimization approaches: code auto-generation (PPCG), vector-level data reuse (Bricks), shared memory optimization (SHM), accumulated summations optimization (SSAM), and temporal blocking (STENCILGEN). For a fair comparison, when comparing with STENCILGEN~\cite{rawat2018domain}, SSAM~\cite{chen2019versatile}, and Bricks~\cite{zhao2019exploiting}, we use the default setting (including the default domain size) in their papers, except that we adjusted the build system to add the latest GPU (A100). We use SSAM's setting to evaluate PPCG~\cite{verdoolaege2013polyhedral}.

We use the test data provided by STENCILGEN~\cite{rawat2018domain}. We tested three PERKS (SHM) implementations: PERKS\_SM that only uses shared memory to cache data; PERKS\_REG that only uses registers to cache data; and PERKS\_MIX that uses both shared memory and registers to cache data. Due to space limitations, we report only the peak performance among those three PERKS variants.
\subsubsection{Conjugate Gradient Datasets}
The datasets for conjugate gradient come from the SuiteSparse Matrix Collection~\cite{davis2011university}. We selected symmetric positive definite matrices that can converge in a CG solver. The details of the selected datasets are listed in Table~\ref{tab:matrxset}. 

We compare the performance of PERKS (CG Solver) with  Ginkgo library~\cite{anzt2020ginkgo}, a widely used library heavily optimized for GPUs (including A100). We run 10,000 time steps in our performance evaluation (similar to Ginkgo's basic setting~\cite{anzt2020ginkgo}). We report the speedup per time step and the measured sustained bandwidth achieved by Ginkgo. For PERKS \revision{(CG Solver)}, we run different variants that implement caching the vector $r$ or the matrix $A$ plus the additional caching of TB-level search result and thread-level search result policies. We only report the best-performing variant for each dataset.% due to space limitations. 
%As mentioned in Section~\ref{sec:impCg}. 

\subsection{\revision{Sizes of Domains and Problems}}\label{sec:finddomain}
\begin{comment}

PERKS intuitively favors small domain/problem sizes. However, for a fair evaluation of PERKS, we can not choose arbitrarily small domain sizes; we need domain/input sizes that fully utilize the compute capability of the device. We conducted an elaborate set of experiments for every individual stencil benchmark to identify the minimum domain size that would fully utilize the device. We use this domain size to represent large domains when PERKS can only partially cache the domain. Note that domain/problem sizes that are beyond domain/problem sizes that could fully utilize the device are effectively serialized by the device once we go beyond peak concurrency sustainable by the device. Table~\ref{tab:stencil_speedups} summarizes the domain sizes (marked as 'P') for stencil benchmarks that would provide a base for a fair comparison. We also test small domains where the whole domain can be cached by PERKS (marked 'F' in Table~\ref{tab:stencil_speedups}).
\end{comment}

PERKS intuitively favors smaller domain/problem sizes. However, for a fair evaluation of PERKS, we can not choose arbitrarily small domain sizes; we need domain/input sizes that fully utilize the compute capability of the device. Similar to~\cite{saturationGPU}, we conducted an elaborate set of experiments for every individual stencil benchmark to identify the minimum domain size that would fully utilize the device. Note that domain/problem sizes that are beyond domain/problem sizes that could fully utilize the device are effectively serialized by the device once we go beyond peak concurrency sustainable by the device. Table~\ref{tab:domain} summarizes the domain sizes for the different stencil benchmarks that would \revision{achieve a fair performance in the SHM implementation}. %\revision{Note that the 3D domains we chose are same lever or larger than the domain size reported by existing research~\cite{gysi2015modesto,li2022efficient,datta2009optimization}}.

For conjugate gradient experiments, we include datasets from SuiteSpare that cover a wide range of problem sizes: from strong-scaling small dataset sizes that would fit in L2 cache and up to large dataset sizes typically reported by libraries for a single GPU of the same generations we use (Gingko~\cite{anzt2020ginkgo,9307857} and MAGMA~\cite{9308884,tnld10}).

\subsection{\revision{Iterative 2D/3D Stencils}}\label{sec:evalstencil}

Figure~\ref{fig:perksresult} compares the performance of SHM~\cite{maruyama2014optimizing} and PERKS based SHM with a wide range of state-of-the-art stencil implementations/libraries. The performance of SHM is comparable to state-of-the-art implementations, i.e. Bricks~\cite{zhao2019exploiting} and SSAM~\cite{chen2019versatile}, across all stencil benchmarks. Applying PERKS consistently speedup SHMs: in comparison to SHM, PERKS (SHM) achieves a the geometric mean speedup of $1.95$x in A100 and $2.31$x in V100, for 2D stencils. The geometric mean speedup for 3D stencils is $1.13$x for A100 and $1.36$x for V100. 

In comparison to the best state-of-the-art spatial blocking implementations/libraries, SSAM, PERKS (SHM) achieves a geometric mean speedup of $1.91$x and $2.11$x for 2D stencils in A100 and V100, respectively. The geometric mean speedup for 3D stencils is $1.15$x for A100 and $1.14$x for V100. In comparison to the state-of-the-art temporal blocking implementation of STENCILGEN, PERKS (SHM) achieves a geometric mean speedup of $1.28$x (A100) and $1.02$x (V100).

\subsection{Conjugate Gradient}\label{sec:evaCG}

Figure~\ref{Fig:cg_compare} compares PERKS (CG Solver) to Ginkgo. \revision{We observe significantly higher performance advantage when the input size is within the L2 cache capacity. This phenomenon implies that PERKS (CG Solver) automatically benefits from the large L2 cache, possibly because the constant matrix can reside in the L2 cache, saving memory traffic to the global memory.} When the input is less than the L2 capacity, PERKS running on A100 achieves a geometric mean of $4.49$x 
speedups; in V100, PERKS achieves $5.50$x speedups. When the input matrix exceeds the L2 cache capacity, PERKS running on A100 achieves a geometric mean of $1.18$x speedup. On V100, PERKS achieves a geometric mean of $1.64$x (double) speedups. It is important to remember that the Ginkgo library that we use as baseline is among the top performing libraries in CG solvers, emphasizing GPU optimizations~\cite{anzt2020ginkgo}.

Note that regardless of whether we stay within the L2 cache capacity or exceed it, we are still caching the domain using one of the caching policies we described earlier. %Yet, upon analyzing the results, we observed a clear drop in speedup once we exceeded the L2 capacity.

{

}

\section{Related Work}\label{sec:related}
% \textit{Memory-bound iterative solvers:} Much research focused on optimizing memory-bound iterative solvers. Some focus on algorithm-level optimizations~\cite{pearson2020preconditioners,anzt2017preconditioned}, 
% some focus on optimizing data access or data reuse at per time step perspective~\cite{Tessellating,10.1145/3174243.3174248,chen2019versatile,zhao2019exploiting,aliaga2015systematic,anzt2020ginkgo}, some conducting temporal blocking~\cite{MURANUSHI20151303,rawat2018domain,DBLP:conf/cgo/MatsumuraZWEM20} for a small amount of time steps. \textbf{i).} However, our research is different from existing methods by considering all time steps together and being general to any iterative methods, unlike temporal blocking, which is only limited to iterative stencil solvers. 

The concept of persistent threads and persistent kernels dates back to the introduction of CUDA~\cite{aila2009understanding,gupta2012study}. The main motivation for persistence at the time was load imbalance issues with the runtime warp scheduler~\cite{aila2009understanding,chen2010dynamic}. Later research focused on using persistent kernels to overcome the kernel invocation overhead (which was high at the time). GPUrdma~\cite{daoud2016gpurdma} and GPU-Ether~\cite{jung2021gpu} expanded on the concept of persistent kernels to reduce the latency of network communication. 

As on-chip \revision{resources} increased, researchers began to capitalize on data reuse in persistent kernel. Most of them focused on specific applications, GPUrdma~\cite{daoud2016gpurdma} proposed to keep the constant matrix in shared memory. Khorasani et al.~\cite{8574555} proposed to keep parameters in registers. 
% Chen et al. developed a flexible systolic array programming model by leveraging the register cache~\cite{chen2019versatile, chen2018efficient}, which they used to improve the performance of CUDA kernels in image processing applications~\cite{ifdk, dfbp}.
Zhu et al.~\cite{zhu2018sparse} proposed a sparse persistent implementation of recurrent neural networks. To our knowledge, this work is the first to propose a generic and methodological blueprint for accelerating memory-bound iterative applications using persistent kernels.

\section{Conclusion}\label{sec:conclusion}
\vspace{-3pt}
We propose a persistent kernel execution \revision{model} for iterative applications. We enhance performance by moving the time loop to the kernel and cache the intermediate output of each time step \revision{with unused on-chip resources}. We show a notable performance improvement for iterative 2D/3D stencils and a conjugate gradient solver for both V100 and A100 over highly optimized baselines.% We further report notably high speedups in small domain/problem sizes, which is beneficial in strong scaling cases. 

%Finally, as the trending of hardware to support larger cache remained, we anticipate an increasing important role of PERKS in future hardware structures. 

%For future work, we plan to extend PERKS to deep learning and apply compressed inference network with PERKS. We also intend to dive into the compiler to clear obstruction of register overuse issue and further provide auto-code generation tools to relieve the burden for programming with PERKS from user side. 
\begin{acks}
This work was supported by JSPS KAKENHI under Grant Numbers JP22H03600 and JP21K17750. This work was supported by JST, PRESTO Grant Number JPMJPR20MA, Japan. This paper is based on results obtained from JPNP20006 project, commissioned by the New Energy and Industrial Technology Development Organization (NEDO).
This manuscript has been co-authored by UT-Battelle, LLC, under contract DE-AC05-00OR22725 with the US Department of Energy (DOE). The publisher acknowledges the US government license to provide public access under the DOE Public Access Plan (\url{http://energy.gov/downloads/doe-public-access-plan}/).
The authors wish to express their sincere appreciation to Jens Domke, Aleksandr Drozd, Emil Vatai and other RIKEN R-CCS colleagues for their invaluable advice and guidance throughout the course of this research. They also wish to thank Dr. Zhao Tuowen from the SambaNova for the helpful discussions. Finally, the first author would also like to express his gratitude to RIKEN R-CCS for offering the opportunity to undertake this research in an intern program. 
\end{acks}
\bibliographystyle{ACM-Reference-Format}
\pagebreak
% \newpage
% \clearpage
\bibliography{acmart}

\end{document}